%% file: ms.tex
\documentclass[12pt,preprint]{aastex}

\tighten

\begin{document}

\title{ Stellar Collisions and the Interior Structure of Blue
Stragglers }

\author{James C.~Lombardi, Jr. and Jessica S.\ Warren\altaffilmark{1}} \affil{Department of Physics and Astronomy,
Vassar College, 124 Raymond Avenue, Poughkeepsie, NY 12604-0745}
\email{lombardi@vassar.edu, jesawyer99@alum.vassar.edu}

\author{Frederic A.~Rasio\altaffilmark{2}} \affil{Department of
Physics, MIT 6-201, Cambridge, MA 02139} \email{rasio@northwestern.edu}

\author{Alison Sills\altaffilmark{3}} \affil{Department of Physics and
Astronomy, Leicester University, England, LE1 7RH}
\email{asills@mcmail.cis.mcmaster.ca}

\and

\author{Aaron R.\ Warren\altaffilmark{1}} \affil{Department of Physics
and Astronomy, Vassar College, 124 Raymond Avenue, Poughkeepsie, NY
12604-0745} \email{aawarren00@alum.vassar.edu}

\altaffiltext{1}{Present address: Dept.\ of Physics and Astronomy,
Rutgers Univ., Piscataway, NJ 08854} \altaffiltext{2}{Present address:
Dept.\ of Physics and Astronomy, Northwestern Univ., Evanston, IL
60208} \altaffiltext{3}{Present address: Dept.\ of Physics and Astronomy, McMaster
Univ., Hamilton, Ontario, L8S 4M1, Canada}

\begin{abstract}
Collisions of main sequence stars occur frequently in dense star
clusters. In open and globular clusters, these collisions produce
merger remnants that may be observed as blue stragglers. Detailed
theoretical models of this process require lengthy hydrodynamic
computations in three dimensions. However, a less computationally
expensive approach, which we present here, is to approximate the
merger process (including shock heating, hydrodynamic mixing, mass
ejection, and angular momentum transfer) with simple algorithms based
on conservation laws and a basic qualitative understanding of the
hydrodynamics. These algorithms have been fine tuned through
comparisons with the results of our previous hydrodynamic simulations.
We find that the thermodynamic and chemical composition profiles of
our simple models agree very well with those from recent SPH (smoothed
particle hydrodynamics) calculations of stellar collisions, and the
subsequent stellar evolution of our simple models also matches closely
that of the more accurate hydrodynamic models. Our algorithms have
been implemented in an easy to use software package, which we are
making publicly available (see
http://vassun.vassar.edu/$\sim$lombardi/mmas/). This software could be
used in combination with realistic dynamical simulations of star
clusters that must take into account stellar collisions.
\end{abstract}

\keywords{celestial mechanics, stellar dynamics -- globular clusters:
general -- hydrodynamics -- stars: blue stragglers -- stars: evolution
-- stars: interiors -- stars: rotation}

%\clearpage
\section{Introduction and Motivation}

Blue stragglers are stars that appear along an extension of the main
sequence (MS), beyond the turnoff point in the color-magnitude diagram
(CMD) of a cluster.  All observations suggest that blue stragglers are
indeed more massive than a turnoff star and are formed by the merger
of two or more parent MS stars.  In particular, \citet{sha97} and
\citet{sep00} have directly measured the masses of several blue
stragglers in the cores of 47 Tuc and NGC 6397 and confirmed that they
are well above the MS turnoff mass, some even with masses apparently
above {\it twice\/} the turnoff mass.  Furthermore, \citet{gil98} have
demonstrated that the masses estimated from the pulsation frequencies
of four oscillating blue stragglers in 47 Tuc are consistent with
their positions in the CMD.

Stellar mergers can occur through either a direct collision or the
coalescence of a binary system \citep{leo89,liv93,str93,bai95}.
Single-single star collisions occur with appreciable frequency only in
the cores of the densest clusters \citep{hil76}, but in lower-density
clusters collisions can also happen indirectly, during resonant
interactions involving binaries
\citep{leo89,leo91,sig94,sig95,dav95,bac96}.  Merger rates depend
directly on cluster properties such as the local density, velocity
dispersion, mass function, and binary fraction.  When mergers do
occur, all of these cluster properties are affected.  The dynamics of
a cluster, including mass segregation and the rate of core collapse,
are consequently influenced, leading to an intricate relation between
individual stellar interactions and cluster evolution
\citep{hut92,ras01}.  By studying stellar mergers, we are therefore
able to probe the dynamics of globular clusters. Results from ongoing
{\it Hubble Space Telescope} surveys of nearby globular clusters
continue to expand the statistics of blue straggler populations,
making it timely for a detailed comparison between observations and
theory.

The final structure and chemical composition profiles of merger
remnants are of central importance, since they determine the
subsequent observable properties and evolutionary tracks of merger
products in a CMD \citep{sil99}. Three-dimensional hydrodynamic
simulations is one means by which we can focus on fluid mixing during
stellar mergers and determine the structure of a remnant.  Many such
simulations of stellar mergers have been presented in the literature
\citep{lom96,san97,sil97,sil01}.  The problem with these simulations,
if they were to be coupled with calculations of the cluster as a
whole, is the prohibitive computing time: high resolution hydrodynamic
simulations can typically take hundreds or even thousands of hours to
complete.

%making it essentially impossible to couple these simulations with
%calculations of a cluster as a whole.

In this paper, we develop a method for computing the structure and
composition profiles of zero-age blue stragglers {\it without} running
hydrodynamic simulations.  Since our method takes considerably less
than a minute to generate a model on a typical workstation, we are
able to explore the results of collisions in a drastically shorter
amount of time. Our approach can be generalized to work for more than
two parent stars, simply by colliding two stars first and then
colliding the remnant with a third parent star. Most importantly, such
algorithms will make it possible to incorporate the effects of
collisions in simulations of globular clusters as a whole.

\section{Procedure}

We begin with two (non-rotating) parent star models, specifying
initial profiles for the stellar density $\rho$, pressure $P$, and
abundance of chemicals as a function of mass fraction.  The profile
for the entropic variable $A\equiv P/\rho^{\Gamma}$, a quantity
closely related (but not equal) to thermodynamic entropy, can also be
calculated easily and is of central importance. Here $\Gamma$ is the
adiabatic index of the gas.  Since the quantity $A$ depends directly
upon the chemical composition and the entropy, it remains constant for
each fluid element in the absence of shocks.

Fluid elements with low values of $A$ sink to the bottom of a
gravitational potential well, and the $A$ profile of a star in stable
dynamical equilibrium increases radially outwards. Indeed, it is
straightforward to show that the condition ${\rm d}A/{\rm d}r>0$ is
equivalent to the usual Ledoux criterion for convective stability of a
nonrotating star \citep{lom96}. The basic idea here can be seen by
considering a small fluid element inside a non-rotating star in
dynamical equilibrium.  If this element is perturbed outward
adiabatically (that is, with constant $A$), then it it will sink back
toward equilibrium only if its new density is larger than that of its
new environment. If instead the element is less dense than its
surroundings, it will continue to float away from the equilibrium, an
unstable situation.  Likewise, if an element is perturbed inward, its
density needs to be less than the environment's in order to return
toward equilibrium.  Since pressure equilibrium between the element
and its immediate environment is established nearly instantaneously,
the ratio of densities satisfies $\rho_{\rm element}/\rho_{\rm
environment}= (A_{\rm element}/A_{\rm environment})^{-1/\Gamma}$, by
the definition of $A$. Therefore if the perturbed element has a larger
$A$ than its new environment, then it has a lower density and buoyancy
will push the element outwards.  Similarly, a fluid element with a
lower $A$ then its surroundings will sink.  As a result, a stable
stratification of fluid requires that the entropic variable $A$
increase outward: ${\rm d}A/{\rm d}r>0$. In such a star, a perturbed
element will experience restoring forces that cause it to oscillate
about its equilibrium position.  For a detailed discussion of the
stability conditions within rotating stars, see \S 7.3 of
\citet{tas78} or \cite{tas00}.  In practice, even in rapidly rotating
stars, fluid distributes itself in such a way that the entropic
variable $A$ increases outwards.

During a collision, the entropic variable $A$ of a fluid element can increase
due to shock heating (see \S\ref{shock}). However, the relative impact
speed of two MS stars in a globular cluster is comparable to the speed
of sound in these parents: both of these speeds are of order
$(GM/R)^{1/2}$, where $G$ is Newton's gravitational constant, and $M$
and $R$ are respectively the mass and length scales of a parent
star. Consequently, the resulting shocks have Mach numbers of order
unity and shock heating is relatively weak. Therefore, to a reasonable
approximation, a fluid element maintains a constant $A$ throughout a
collision.

The underlying principle behind our method exploits the two special
properties of $A$ that were just discussed.  Namely, the entropic
variable $A$ will (1) increase outward in a stable star and (2) be
nearly conserved during a collision.  Therefore, to a good
approximation, {\it the distribution of fluid in a collisional remnant
can be determined simply by sorting the fluid from both parent stars
in order of increasing $A$}: the lowest $A$ fluid from the parent
stars is placed at the core of the remnant and is surrounded by shells
with increasingly higher $A$. In this paper, we will further improve
upon this approximation by also modeling mass loss, shock heating,
fluid mixing, and the angular momentum distribution.

\begin{figure}
\plotone{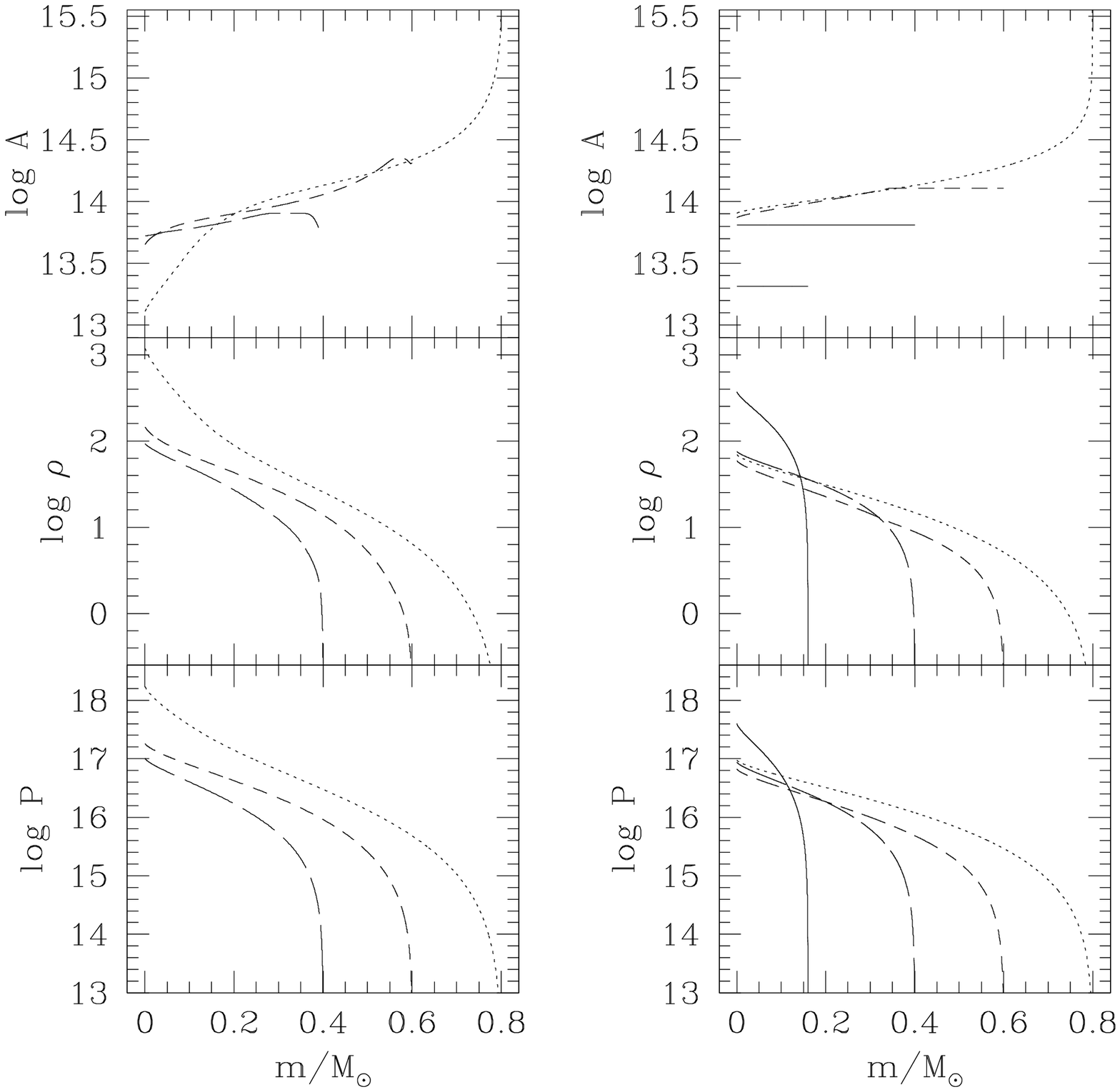}
%\plotone{logthermpar.eps}
\caption{Thermodynamic profiles of
$A(=P/\rho^{5/3})$, pressure $P$, and density $\rho$ as a
function of enclosed mass $m$.
Our three realistically modeled
parent stars are displayed in the left column, and the four polytropic
models are displayed on the right.
The dotted, short-dashed, long-dashed and solid curves
refer to $0.8M_\odot$, $0.6M_\odot$, $0.4M_\odot$ and $0.16M_\odot$ parent
stars, respectively. Logarithms are base 10 and units are cgs.
\label{logthermpar} }
\end{figure}

Our algorithms are calibrated from the results of smoothed particle
hydrodynamic (SPH) calculations presented in \citet{lom96} (for
collisions of polytropic stars) as well as in \citet{sil97} and
\citet{sil01} (for collisions of more realistically modeled stars).
For details and tests of our SPH code, see \citet{lom99}.  For reviews
of SPH, see \citet{mon92} or \citet{ras99}.  Characteristics of the
various parent stars used in our calculations are summarized in Table
\ref{tbl-1}, with thermodynamic profiles shown in Figure
\ref{logthermpar}.  The various collision scenarios we have considered
are listed, along with mass loss information, in Table \ref{tbl-2}.

\begin{figure}
\plotone{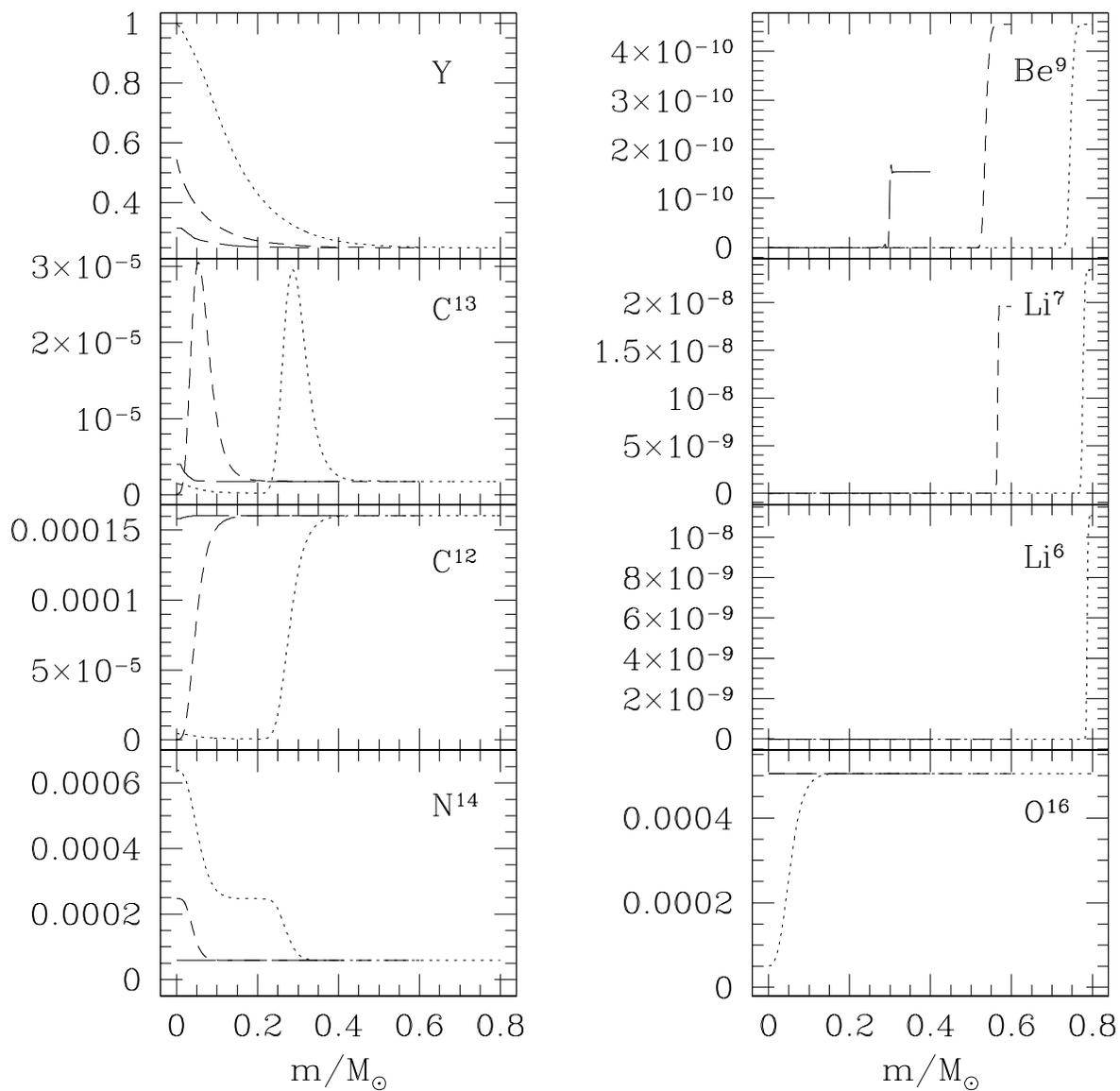}
%\plotone{chemparmdl.eps}
\caption{Fractional chemical abundance (by mass)
as a function of enclosed mass $m$ for various chemical elements
in our three realistically modelled parent stars.  Line types are as in Fig.\
\ref{logthermpar}. \label{chempar} }
\end{figure}

The realistically modeled parent stars are based on calculations
done with the Yale Rotating Evolution Code (YREC), as discussed in
\citet{sil97}.  In particular, we evolved non-rotating
MS stars with a primordial helium abundance $Y=0.25$
and metallicity $Z=0.001$ for 15 Gyr, the amount of time needed to
exhaust the hydrogen in the center of the $0.8M_\odot$ star.  We
note that $P/\rho^{5/3}$ decreases slightly in the outermost
layers of the $0.4M_\odot$ and $0.6M_\odot$ stars modeled by YREC (see Fig.\
\ref{logthermpar}).  The adiabatic index $\Gamma$ is actually less
than the ideal gas value of $5/3$ in these regions, due to the
relative importance of ionization and radiation pressure.  In this
paper, however, we neglect these effects and instead simply force
the $A$ profile to increase by some negligibly small amount in
these regions. Figure \ref{chempar} displays chemical
abundance profiles of these parent stars.

\subsection{Mass Loss} \label{massloss}

The velocity dispersion of globular cluster stars is typically
only $\sim 10$ km s$^{-1}$, which is much smaller than the escape
velocity from the surface of a MS star; for example, a star of
mass $M=0.8 M_\odot$ and radius $R=R_\odot$ has an escape velocity
$(2GM/R)^{1/2}=552$ km s$^{-1}$.  For this reason, collisional
trajectories are well approximated as parabolic, and the mergers
are relatively gentle: the mass lost is never more than about 8\%
of the total mass in the system [mass loss with hyperbolic
trajectories is treated by \citet{lai94}].  Furthermore, most MS stars in globular clusters are not rapidly rotating, and
it is a good approximation to treat the initial parent stars as
non-rotating.

Given models for the parent stars (see Table \ref{tbl-1}), we first
determine the mass lost during a collision.  Inspection of
hydrodynamic results for collisions between realistically modeled
stars, as well as for collisions between polytropes, suggests that the
fraction of mass ejected can be estimated approximately by
\begin{equation}
f_L = c_1 {\mu \over M_1+M_2} {R_{1,0.86} +R_{2,0.86} \over R_{1,0.5}
+ R_{2,0.5} +c_2 r_p}, \label{fL}
\end{equation}
where $c_1$ and $c_2$ are dimensionless constants that we take to be
$c_1=0.157$ and $c_2=1.8$, $\mu\equiv M_1M_2/(M_1+M_2)$ is the reduced
mass of the parent stars, $R_{i,0.5}$ and $R_{i,0.86}$ are the radii
in parent star $i$ enclosing a mass fraction $m/M_i=0.5$ and 0.86,
respectively, and $r_p$ is the periastron separation for the initial
parabolic orbit. While developing equation (\ref{fL}) we searched for
a relation that accounted for the {\it mass distribution} (not just
the total masses and radii) of the parent stars in some simple
way. The more diffuse the outer layers of the parents, the longer the
stellar cores are able to accelerate toward each other after the
initial impact: the sum of half-mass radii, $R_{1,0.5} + R_{2,0.5}$, in the denominator of
equation (\ref{fL}) accounts for this increased effective collision
speed for parents whose mass distributions are more centrally
concentrated.  The dependence on $\mu$ in equation (\ref{fL}) arises
from the expectation that the mass loss will be roughly proportional
to the kinetic energy at impact, and from the fact that a simple
rescaling of the stellar masses ($M_i \rightarrow k M_i$) in a
hydrodynamic simulation leaves $f_L$ unchanged.

The final (post-shock) value of a fluid element's entropic variable
will be larger than its initial value, as discussed in \S\ref{shock}.
The mass loss must be distributed between the two parent stars such
that the outermost fluid layers retained from each parent has the same
final entropic variable $A$, so that the layers can merge together
into a stable equilibrium.  We solve for this maximum value of $A$ in
the remnant by requiring that the mass of the fluid with larger final
$A$ be the desired ejecta mass.  This constraint determines what
fraction of the ejecta comes from each of the parent stars.

\subsection{Shock Heating} \label{shock}

Shocks increase the value of a fluid element's entropic variable
$A=P/\rho^{\Gamma}$.  The distribution and timing of shock heating
during a collision involve numerous complicated processes: each impact
generates a recoil shock at the interface between the stars, the
oscillating merger remnant sends out waves of shock rings, and finally
the outer layers of the remnant are shocked as gravitationally bound
ejecta fall back to the remnant surface. For off-axis collisions this
may be repeated several times.  Our goal is not to {\it derive}
approximations describing the shock heating during each of these
stages, but rather empirically to determine physically reasonable
relations that fit the available SPH data.

%Nevertheless, because the impact speed for these
%parabolic collisions is on the same order of magnitude as the sound
%speed, large Mach numbers are never achieved and shock heating is never
%extrodinarily large.

Let $A$ and $A_{init}$ be, respectively, the final and initial values
of the entropic variable for some particular fluid element.  We used
the results of hydrodynamic calculations to examine how the change
$A-A_{init}$, as well as the ratio $A/A_{init}$, depended on a variety
of functions of $P_{init}$ (the initial pressure) and $A_{init}$.  Our
search for a simple means of modeling this dependence was guided by a
handful of features evident from hydrodynamic simulations: (1) fluid
deep within the parents are shielded from the brunt of the shocks, (2)
in head-on collisions, fluid from the less massive parent experiences
less shock heating than fluid with the same initial pressure from the
more massive parent, (3) in off-axis collisions with multiple
periastron passages before merger, fluid from the less massive parent
experiences more shock heating than fluid with the same initial
pressure from the more massive parent, and (4) the shock heating
within each parent clearly must be the same if the two parent stars
are identical.  In all of the hydrodynamic calculations considered we
model the fluid in our system using an adiabatic index $\Gamma=5/3$,
corresponding to an ideal gas equation of state.

%The entropic variable profile of the remnant, as a function of enclosed
%mass, depends weakly on the periastron separation $r_p$ (see
%Fig.\ 13 of \citet{lom96}).

\begin{figure}
\plotone{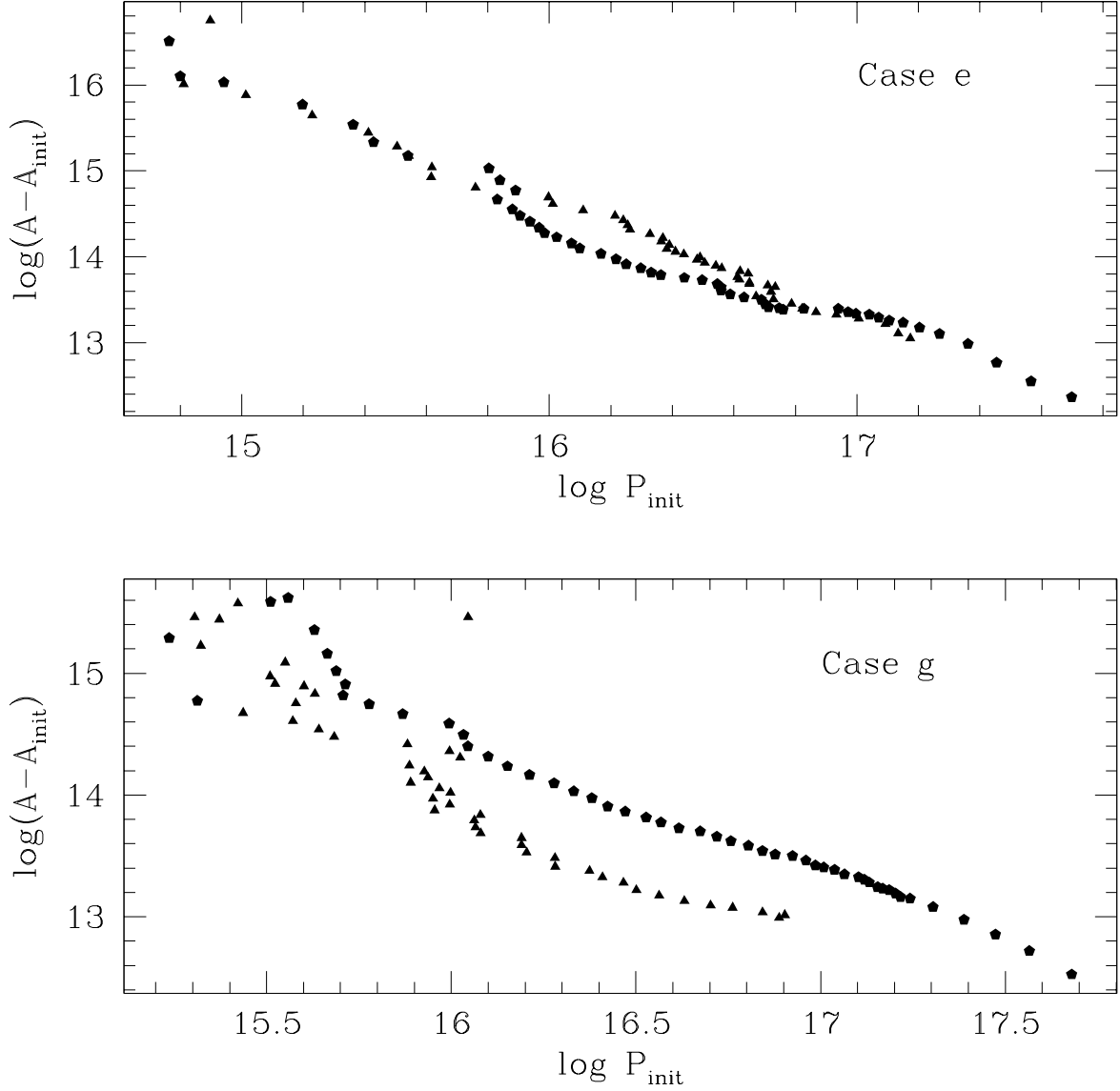}
%\plotone{eglogp.eps}
\caption{The change in entropic variable $A$ as a
function of initial pressure $P_{init}$ on a log plot for the SPH
remnant of Case e and Case g (see Table \ref{tbl-2} for data describing these cases).  Pentagons
refer to fluid from parent star~1 that has reached dynamical
equilibrium by the end of the simulation and that has been binned
by enclosed mass fraction; triangles refer to the corresponding
fluid from star~2.
Logarithms are base 10 and units are cgs. \label{eglogp} }
\end{figure}

We find that when $\log_{10}(A-A_{init})$ is plotted versus
$\log_{10}P_{init}$, the resulting curve for each parent star is
fairly linear (see Fig.\ \ref{eglogp}) with a slope of approximately
$c_3=-1.1$ throughout most of the remnant in the $\sim 25$ simulations
we examined:
\begin{equation}
\log_{10} (A-A_{init}) = b_i(r_p)+c_3\log_{10} P_{init}, ~~~ i=1,2
\label{delA}.
\end{equation}
Here the intercept $b_i(r_p)$ is a function of the periastron
separation $r_p$ for the initial parabolic trajectory as well as the
masses $M_1$ and $M_2$ of the parent stars.  Larger values of $b_i$
correspond to larger amounts of shock heating in star $i$, where the
index $i=1$ for the more massive parent and $i=2$ for the less massive
parent ($M_2<M_1$).  For simplicity of notation, we have suppressed
the index $i$ on the $A$, $A_{init}$ and $P_{init}$ in equation
(\ref{delA}).

The SPH data suggest that the intercepts $b_i(r_p)$ can be fit
according to the relations
\begin{eqnarray}
b_1(r_p) & = & b_1(0) - c_4 {r_p \over R_1+R_2} \log_{10}(M_1/M_2)
\label{b1}\\ b_2(r_p) & = & b_1(0) + \left(c_5 {r_p \over R_1+R_2}
-c_6\right) \log_{10}(M_1/M_2), \label{b2}
\end{eqnarray}
where $c_4=0.5$, $c_5=5$, $c_6=2.5$, and $b_1(0)$ is the intercept for
a head-on collision ($r_p=0$) between the two parent stars under
consideration.

Although equations (\ref{delA}), (\ref{b1}) and (\ref{b2}) describe
how to distribute the shock heating, the overall strength of the shock
heating hinges on the value chosen for $b_1(0)$.  To determine
$b_1(0)$, we consider the head-on collision between the parent stars
under consideration and exploit conservation of energy: more
specifically, we choose the value of $b_1(0)$ that ensures that the
initial energy of the system equals the final energy during a head-on
collision. Since we are considering parabolic collisions, the orbital
energy is zero and the initial energy is simply $E_{tot}=E_1+E_2$, the
sum of the energies for each of the parent stars.  The final energy of
the system includes energy associated with the ejecta and the center
of mass motion of the remnant, in addition to the energy $E_r$ of the
remnant in its own center of mass frame.  In this paper we consider
non-rotating parent stars, and so the remnant of a head-on collision
also is non-rotating and its structure quickly approaches spherical
symmetry.  The values of $E_1$, $E_2$ and $E_r$ are therefore simply
the sum of the internal and self-gravitational energies calculated
while integrating the equation of hydrostatic equilibrium. Since the
energy $E_r$ depends on the thermodynamic profiles of the remnant, it
is therefore a function of a shock heating parameter $b_1(0)$ (see \S
\ref{mixing} for the details of how the remnant's structure is
determined).

The energy $E_r$ of the remnant is nearly equal to the initial energy
$E_{tot}$ of the system.  However, the ejecta do carry away a portion
of the total energy, suggesting that the energy conservation equation
be written as
\begin{equation}
E_{tot}=E_r-c_7 f_L E_{tot}, \label{Etot}
\end{equation}
where the coefficient $c_7$ is order unity and $f_L$ is the fraction
of mass lost during the collision (see \S\ref{massloss}).  We use a
value of $c_7=2.5$, which is consistent with the available SPH
data (see Table \ref{tbl-energy}). In equation (\ref{Etot}), the left
hand side is the initial energy of the system, and the right hand side
is its final energy.  The second term on the right hand side accounts
for the energy associated with the ejecta and with any center of mass
motion of the remnant (note that this term is positive since
$E_{tot}<0$). In practice, we iterate over $b_1(0)$ until equation
(\ref{Etot}) is solved.  Equation (\ref{Etot}) needs to be solved only
once for each pair of parent star masses $M_1$ and $M_2$: once
$b_1(0)$ is known, we can model shock heating in a collision with any
periastron separation $r_p$ by first calculating $b_1(r_p)$ and
$b_2(r_p)$ from equations (\ref{b1}) and (\ref{b2}) and by then using
these values in equation (\ref{delA}).

\subsection{Merging and Fluid Mixing}\label{mixing}

As with any star in stable dynamical equilibrium, the remnant will
have an $A$ profile that increases outward.  In our model, fluid
elements with a particular post-shock $A$ value in both parent stars will merge to
become the fluid in the remnant with the same value of the entropic
variable.  Furthermore, if the fluid in the core of one parent star
has a lower $A$ value than any of the fluid in the other parent star,
the former's core must become the core of the remnant, since the
latter cannot contribute at such low entropies.  When merging the
fluid in the two parent stars to form the remnant, we use the post
shock entropic variable $A$, as determined from equation (\ref{delA}).

Within the merger remnant, the mass $m_r$ enclosed within a surface of
constant $A$ must equal the sum of the corresponding enclosed masses
in the parents:
\begin{equation}
\left. m_r\right|_{A_r=A} = \left. m_1\right|_{A_1=A} +
\left. m_2\right|_{A_2=A}.
\end{equation}
%
%$m_r(A_r=A)=m_1(A_1=A)+m_2(A_2=A)$,
%
It immediately follows that the derivative of the mass in the remnant
with respect to $A$ equals the sum of the corresponding derivatives in
the parents: d$m_r/$d$A_r=$d$m_1/$d$A_1+$d$m_2/$d$A_2$, or $dA_r/dm_r
= [(dA_1/dm_1)^{-1}+(dA_2/dm_2)^{-1}]^{-1}$.  In practice, we calculate these
derivatives using simple finite differencing.  If we
partition the parent stars and merger remnant into mass shells, then two
adjacent shells in the remnant that have enclosed masses that differ
by $\Delta m_r$ will have entropic variables that differ by
\begin{equation}
\Delta A_r = {\Delta m_r \over \left({d A_1\over dm_1}\right)^{-1} +
\left({dA_2\over dm_2}\right)^{-1}}. \label{DeltaAr}
\end{equation}
The value of $A$ at a particular mass shell in the remnant is then
determined by adding $\Delta A_r$ to the value of $A$ in the previous
mass shell.

In the case of the (non-rotating) remnants formed in head-on
collisions, knowledge of the $A$ profile is sufficient to determine
uniquely the pressure $P$, density $\rho$, and radius $r$ profiles.
While forcing the $A$ profile to remain as was determined from sorting
the shocked fluid, we integrate numerically the equation of
hydrostatic equilibrium with d$m=4\pi r^2\rho$d$r$ to determine the
$\rho$ and $P$ profiles [which are related through
$\rho=(A/P)^{3/5}$]. This integration is an iterative process, as we
must initially guess at the central pressure.  Our boundary condition
is that the pressure must be zero when the enclosed mass equals the
desired remnant mass $M_r=(1-f_L)(M_1+M_2)$.  During this numerical
integration we also determine the remnant's total energy $E_r$ and
check that the virial theorem is satisfied to high accuracy. The total
remnant energy $E_r$ appears in equation (\ref{Etot}), and if this
equation is not satisfied to the desired level of accuracy, we adjust
our value of $b_1(0)$ accordingly and redo the shocking and merging
process.

As done in \citet{sil01}, the structure of a rotating remnant can be
determined by integrating modified equations of equilibrium [see eq.\
(9) of \citet{ES76}], once the entropic variable $A$ and specific angular
momentum $j$ distribution are known (see \S\ref{angular}). To do so,
one can implement an iterative procedure in which initial guesses at
the central pressure and angular velocity are refined until a
self-consistent model is converged upon.  Even for the case of
off-axis collisions and rotating remnants, the chemical composition
profiles can still be determined, even without solving for the
pressure and density profiles, as we will now discuss.

Once the $A$ profile of the remnant has been determined, we focus our
attention on its chemical abundance profiles.  Not all fluid with the same
initial value of $A_{init}$ is shock heated by the same amount during
a collision, since, for example, fluid on the leading edge of a parent
star is typically heated more violently than fluid on the trailing
edge of the parent.  Consequently, fluid from a range of initial
shells in the parents can contribute to a single shell in the remnant.
To model this effect, we first mix each parent star by spreading out
its chemicals over neighboring mass shells, using a Gaussian-like
distribution that depends on the difference in enclosed mass between
shells.  Let $X_i$ be the chemical mass fraction of some species $X$
in a particular shell $i$, and let the superscripts ``pre'' and
``post'' indicate pre- and post-mixing values, respectively.  Then
\begin{eqnarray}
X_k^{post} & = & \sum_i{X_i^{pre} g_{ik} \Delta m_i \over \sum_j
g_{ji} \Delta m_j}, \label{Xkpost} \\
g_{ik} & = & \exp\left[-{\alpha
\over M^2}(m_i-m_k)^2\right]+\exp\left[-{\alpha \over
M^2}(m_i+m_k)^2\right]+\exp\left[-{\alpha \over
M^2}(m_i+m_k-2M)^2\right], \label{gik}\\
\alpha & = & c_8 \left[ \ln\left( A_{max}/A_{min} \right) \right]^2, \label{alpha}
\end{eqnarray}
where $\Delta m_i$ is the mass of shell $i$, $m_i$ is the mass
enclosed by shell $i$, $M$ is the total mass of the parent star,
$c_8$ is a dimensionless coefficient that we choose to be $c_8=5$, and
$A_{max}$ and $A_{min}$ are the maximum and minimum post-shock entropic
variables of fluid that will be gravitationally bound to the remnant.
We have suppressed an additional index in equations (\ref{Xkpost})
through (\ref{alpha}) that would label the parent star. The summand in
equation (\ref{Xkpost}) is the contribution from shell $i$ to shell
$k$.  The second term in the distribution function, equation
(\ref{gik}), is important only for mass shells near the center of the
parent, while the third term becomes important only for mass shells
near the surface; these two correction terms guarantee that an
initially chemically homogeneous star remains chemically homogeneous
during this mixing process ($X_k^{post}=X_k^{pre}=$constant, for any
shell $k$).  The dependence of $\alpha$ on $A_{max}/A_{min}$ ensures that stars with steep entropy gradients are more
difficult to mix [see Table 4 of \citet{lom96}].

Consider a fluid layer of mass $dm_r$ in the merger remnant with a post-shock
entropic variable $A$ that ranges from $A_r$ to $A_r+dA_r$.  The
fraction of that fluid $dm_i/dm_r$ that originated in parent star $i$
can be calculated as $(dA_r/dm_r)/(dA_i/dm_i)$.  Therefore, the
composition of this fluid element can be determined by the weighted
average
\begin{equation}
X_r = X^{post,1} {dA_r/dm_r \over dA_1/dm_1} + X^{post,2} {dA_r/dm_r
\over dA_2/dm_2}, \label{xr}
\end{equation}
where all derivatives are evaluated at $A_r$, the post-shock value of the
entropic variable under consideration.
With the post-shock $A$ profiles given by equation (\ref{delA}) and the smoothed composition profiles given by equation (\ref{Xkpost}),
equation (\ref{xr}) allows us to merge the parent stars and determine the final composition profile of the remnant.

\subsection{Angular Momentum Distribution}\label{angular}

To estimate the total angular momentum $J_r$ of the remnant in its
center of mass frame, we use angular momentum conservation in the same
way that energy conservation was used in \S\ref{shock}.  In
particular, since the parent stars are non-rotating, the total angular
momentum in the system is just the orbital angular momentum,
\begin{equation}
J_{tot}=M_1M_2\left({2 G r_p\over M_1+M_2}\right)^{1/2},
\end{equation}
which we set equal to $J_r$ plus a contribution due to mass loss [cf.\
eq.\ (\ref{Etot})]:
\begin{equation}
J_{tot}=J_r+c_9 f_L J_{tot}. \label{Jtot}
\end{equation}
The SPH simulations demonstrate that $J_{tot}$ is always slightly
larger than $J_r$, and the choice $c_9=2$ leads to good agreement with
the SPH results.  Equation (\ref{Jtot}) can be solved for $J_r$, and
the results are compared with those of SPH simulations
in Table \ref{tbl-angular_momentum}.  The agreement
(between the numbers in the last two columns) is excellent for all
cases with $M_1/M_2 \le 2$.  For Cases V and W, which have a
relatively large mass ratio ($M_1/M_2=5$), the approximation begins to
falter, although the predicted $J_r$ value still agrees with SPH
results to within 10\%.

The structure of the rotating remnants formed in off-axis
collisions depends on the distribution of the specific angular
momentum within the remnant.  Even though the collisional remnants
are axisymmetric around the rotation axis with angular velocities
$\Omega$ that are constant on isodensity surfaces, the specific
angular momentum distribution can nevertheless be quite
complicated [see Fig.\ 12 of \citet{lom96} or Fig.\ 3 of
\citet{sil01}].  The goal here is to simplify this complicated
distribution into an average one-dimensional profile.  The
specific angular momentum $j$ for the SPH remnants increases
outward and is typically concave upward throughout most of the
remnant when averaged over isodensity surfaces and plotted against
enclosed mass.

Once an approximate analytic {\it form} for the average $j$
profile is specified, the profile can be constrained to satisfy
\begin{equation}
J_r=\int_0^{M_r} j(m) dm, \label{Jr}
\end{equation}
where $m$ corresponds to the mass enclosed within a constant
density surface and $J_r$ is determined from equation
(\ref{Jtot}).  We find that the relation
\begin{equation}
j(m)= 
\left\{
  \begin{array}
    {r@{\quad\rm{if}\quad}l}
    c_{10}c_s r(m/ M_r)^{1/3} & m<k_1 M_r, \\
    k_2 \left(G m r\right)^{1/2}(m/M_r)^{1/3} + k_3 & m \ge k_1 M_r,
  \end{array}
\right. \label{jofm}
\end{equation}
with $c_{10}=0.6$, is able to reproduce the important
features of the specific angular momentum profile.  Here, $M_r$ is
the remnant mass, $c_s=(\Gamma P/\rho)^{1/2}$ is the local sound
speed, $r$ is the local radius, and $G$ is Newton's gravitational
constant.  Other forms for $j(m)$ could also be used, and
normalized through equation (\ref{Jr}). One advantage of equation
(\ref{jofm}) is that for $m$ near 0 the specific angular momentum
$j(m)$ scales like $m^{2/3}$, in agreement with both simple
analytic treatments and SPH results. Equation (\ref{jofm}) is
chosen because the rotation in the innermost regions ($m<k_1 M_r$)
of remnants is not strongly affected by $r_p$ and because of the
equation's loose resemblance to the $j$ profile of a Keplerian
disk for $m$ close to $M_r$.  The $c_s$ and $r$ profiles used in
equation (\ref{jofm}) are evaluated for a non-rotating equilibrium
star with the same $A$ profile as the star under consideration, a
simplification that both eases and quickens the necessary
computations.  The coefficients $k_1$, $k_2$ and $k_3$ are not
free parameters, but instead are determined by equation (\ref{Jr})
and by the two additional constraints that $j(m)$ and its
derivative be continuous at $m=k_1 M_r$.  We always choose the
smallest positive value of $k_1$ that meets these constraints.  If no
solution with $k_1>0$ exists, we set $k_1=k_3=0$ 
and solve for $k_2$.  For a non-rotating star clearly $k_1=k_2=k_3=0$.

\section{Results}

\subsection{Comparison with Three-Dimensional Hydrodynamic Calculations\label{comparison}}

To test further the accuracy of our simple models, we compare their
structure and composition against models generated directly from the
results of SPH calculations.  These SPH calculations include both
collisions between polytropic parent models (referred to with a
capital letter as the case name) and collisions between realistically
modeled parent stars (referred to with a lower case letter).

\subsubsection{Shock Heating\label{results:shock}}

\begin{figure}
\plotone{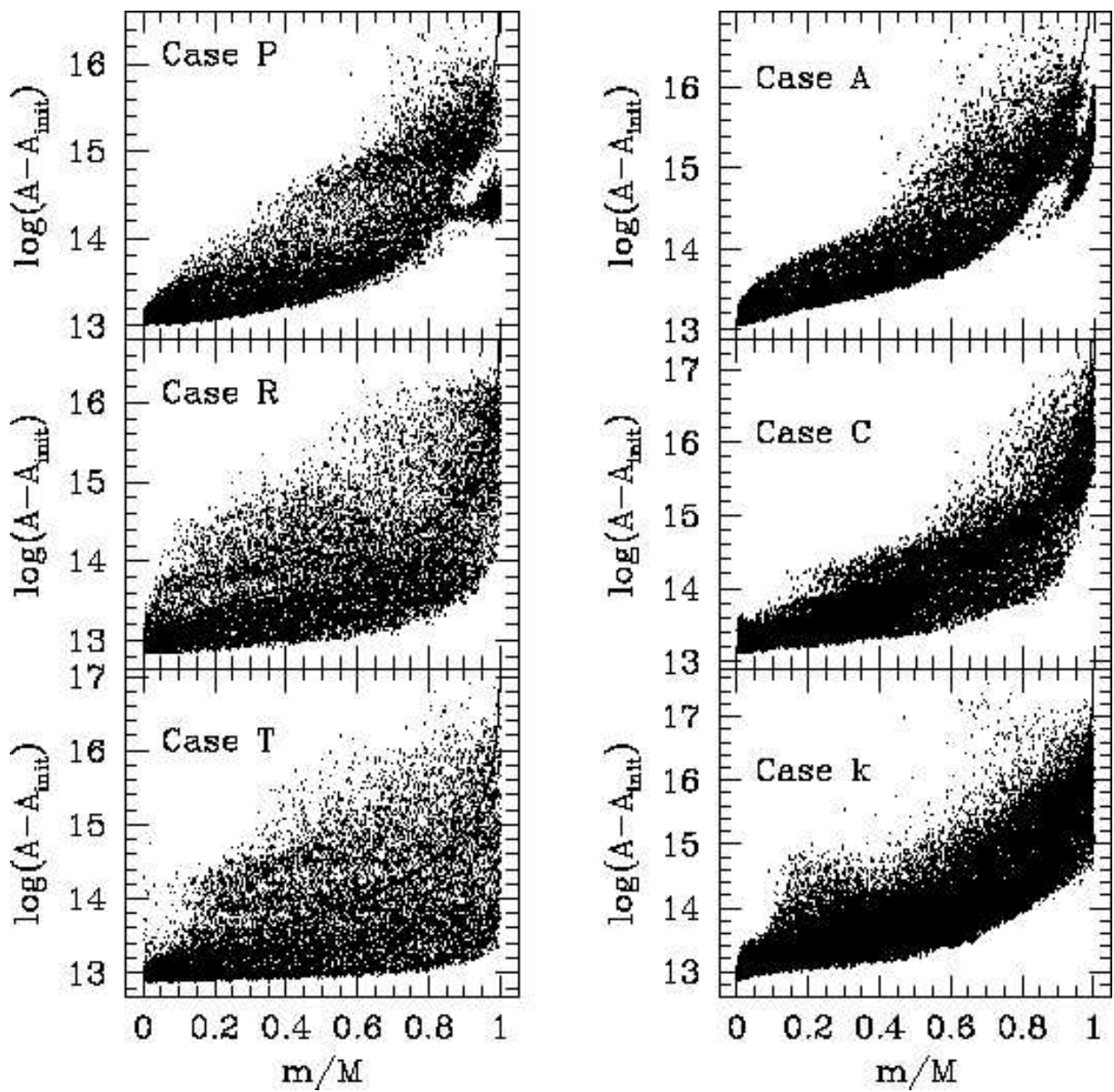}
%\plotone{shockeq.eps}
\caption{
Change in the entropic variable $A$ plotted against the initial enclosed mass fraction $m/M$ for collisions involving identical parent stars.
Each point represents an individual SPH particle, and the solid curve gives the typical increase in $A$ predicted by equation (\ref{delA}).
Logarithms are base 10 and units are cgs. \label{shockeq} }
\end{figure}

\begin{figure}
\plotone{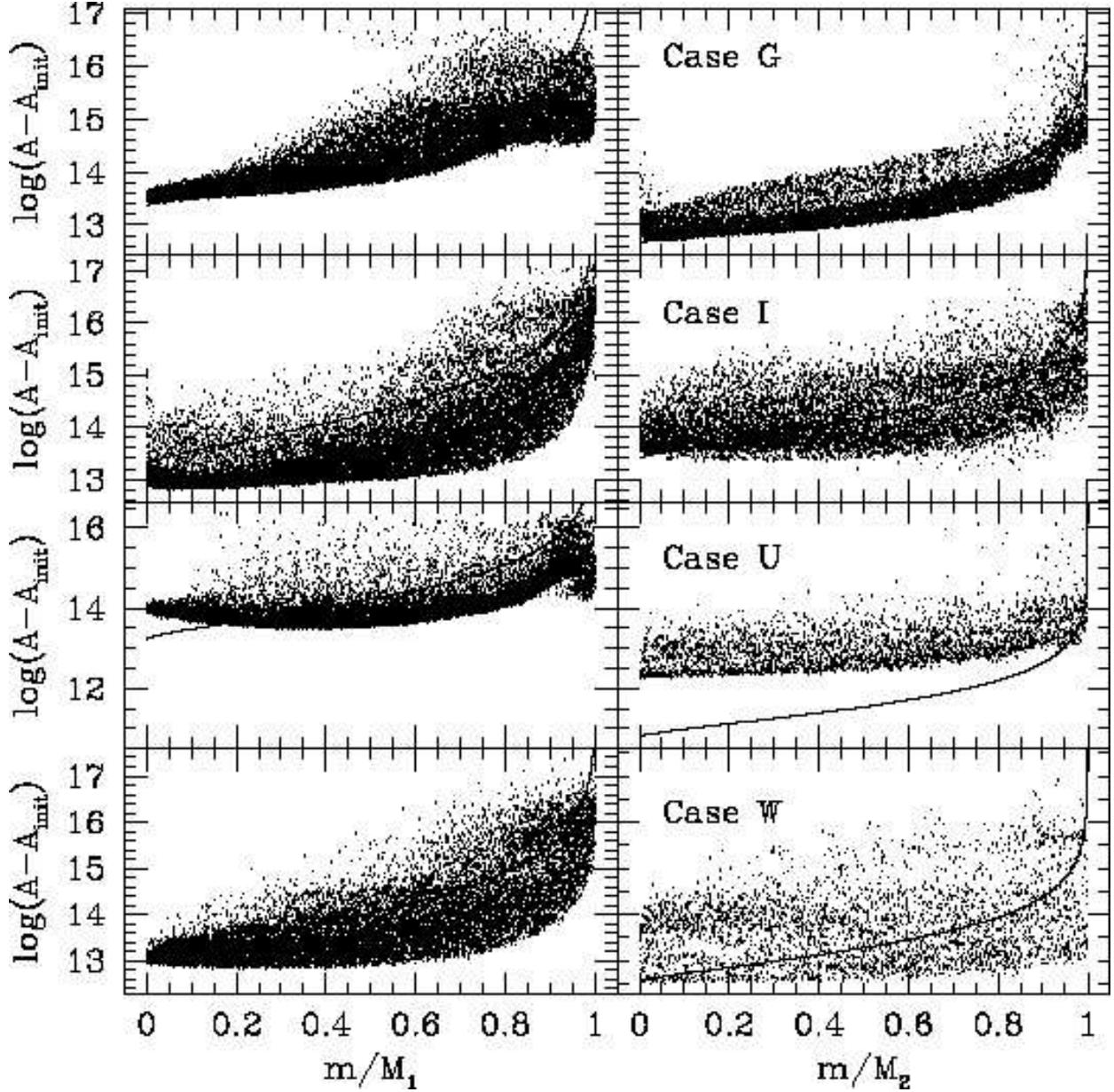}
%\plotone{sGIUW.eps}
\caption{
Change in the entropic variable $A$ plotted against the initial enclosed mass fraction $m/M$ for collisions involving parent stars of unequal mass.
For each collision scenario, the left and right column displays data for the larger and the smaller star, respectively.
Each point represents an individual SPH particle, and the solid curve gives the typical increase in $A$ predicted by equation (\ref{delA}).
Logarithms are base 10 and units are cgs. \label{sGIUW} }

\end{figure}

Clearly, expressions for describing shock heating such as equations
(\ref{delA}), (\ref{b1}) and (\ref{b2}) are rather crude
approximations that lump together complicated effects from the various
stages of the fluid dynamics.  However, these expressions do work
quite well for parent stars of similar mass.  To demonstrate this
point, Figure \ref{shockeq} compares the shock heating of this method
against the heating experienced by the individual SPH particles in six
different collisions of identical parent stars, while Figure
\ref{sGIUW} presents similar data for four collisions between unequal
mass parents.  The agreement between prediction and simulation is
excellent for mass ratios $M_1/M_2$ from 1 to approximately 2,
regardless of the periastron separation $r_p$.  Even for a mass ratio as
large as 5, the prescription continues to work well, at least for
intermediate values of the periastron separation $r_p$ (see Case W in
Fig.\ \ref{sGIUW}).  For head-on collisions with large mass ratios, the
predicted shock heating is an underestimate in the smaller star and in
the center of the larger star (see Case U in Fig.\ \ref{sGIUW}).  It
is worth noting that in collisions between two MS stars, a
remnant must be relatively far past the turnoff on a
CMD if it is to be identified as a blue
straggler.  It is therefore unlikely that collisions involving parent
stars with a mass ratio more than 5 will produce a true blue
straggler, simply because the remnant mass could not be considerably
more than the turnoff mass.

The SPH data in Figures \ref{shockeq} and \ref{sGIUW} clearly shows
that fluid from the same initial enclosed mass fraction $m/M$ can be
shock heated different amounts.  This effect can be easily understood
since, for example, fluid on the impact side of a star will be heated
more than fluid with the same $m/M$ on the back side.  The treatment
of mixing in \S\ref{mixing} does allow us to model this spread in
shock heating, by redistributing fluid to positions with slightly
higher or lower final $A$ values than what is given by equation
(\ref{delA}).  The extent of the redistribution is set by the smoothing
parameter $\alpha$ [see eq.\ (\ref{alpha})], taken to be constant over
each parent star.
Larger values of
$\alpha$ correspond to a smaller width of mass fractions over which
the fluid is distributed.
The parameter $\alpha=248$, for example, for the parents in
Case~a, while for Case~g, $\alpha_1=229$ for the $0.8M_\odot$ parent
and $\alpha_2=143$ for the $0.4M_\odot$ parent.
This approach does well at mimicking the
overall effects of the spread in shock heating.  However, mixing of
fluid is somewhat overestimated in the core and underestimated in the
outer layers, affecting predictions for the central concentration of
helium (see \S\ref{structure}) and for the surface concentration of
trace elements such as lithium (see \S\ref{mass}).  Future treatments
could perhaps improve results by implementing a position dependent
$\alpha$.

Note that our shock heating prescription does necessarily imply the
desirable qualitative features that are discussed in \S\ref{shock} and
evident in the SPH data of Figures \ref{shockeq} and \ref{sGIUW}: (1)
fluid with large initial pressure $P_{init}$ (the fluid shielded by
the outer layers of the star) is generally shock heated less, (2)
$b_2(0)\le b_1(0)$, so that a less massive star is shock heated less
in head-on collisions, (3) $b_1(r_p)$ decreases with $r_p$ while
$b_2(r_p)$ increases with $r_p$, so that for sufficiently large $r_p$
we have $b_2(r_p)>b_1(r_p)$ and the less massive star is shock heated
more, and (4) $b_1(r_p)=b_2(r_p)$ whenever $M_1=M_2$, so that
identical parent stars always experience the same level of shock
heating.

\subsubsection{Mass Loss\label{mass}}

Although mass loss is very small in a parabolic collision, it is
nevertheless important to understand well if one is interested in
tracking trace elements that may only be found in outermost layers of
the parent stars.
The method of \S\ref{massloss} yields
remnant masses that are seldom more than $\sim0.01M_\odot$ different
than what is given by a hydrodynamic simulation (see the last two
columns of Table \ref{tbl-2}); this is clearly a significant
improvement over neglecting mass loss completely, which sometimes can
overestimate the remnant mass by more than $\sim0.1 M_\odot$.
Table \ref{tbl-2} also lists the
fraction $f_2$ of the ejecta that originated in star 2 for each
collision scenario, both as calculated from
our simple method and as calculated from an SPH simulation.
Our simple method allows for the determination of $f_2$ simply by
requiring that the outermost fluid retained from each of the parent
stars has the same post-shock entropy.
Again, the agreement between prediction and simulation is excellent,
especially for cases with mass ratios $M_1/M_2 \lesssim 2$.

\begin{figure}
\plotone{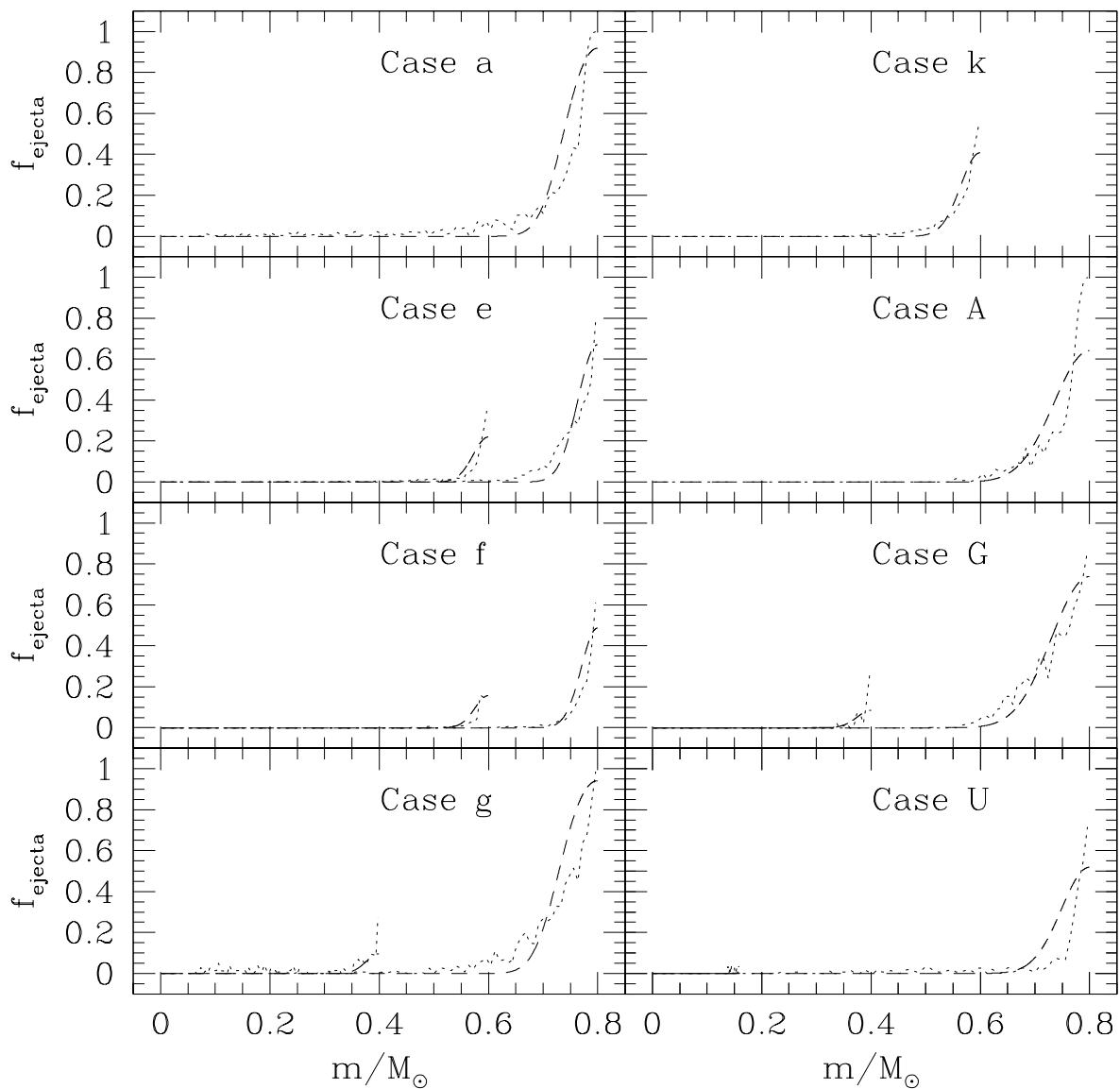}
%\plotone{fe.eps}
\caption{Local fraction of mass that is ejected
by the collision, as a function of the initial enclosed mass $m$
inside each parent star.  The dotted curves represent the results of a 3D SPH simulation, while the dashed curves represent the method of this paper. For Cases a, k and A, only a single
curve is necessary for each line type, since the parent stars are
identical and experience the same mass loss distribution.  \label{fejecta} }
\end{figure}

The mass loss and merging procedures implemented in this paper
allow one to identify not only how much of the ejecta
originated in each parent star, but also the
original location of the fluid within each parent.  The expression
$g_{ik}\Delta m_i \Delta m_k/(\sum_j g_{ji}\Delta m_j)$
gives the amount of mass in shell $i$ that is
transported to shell $k$ [cf.\ eq.\ (\ref{Xkpost})].  Therefore the fraction of the mass from shell $i$
that ultimately becomes ejecta is given by
\begin{equation}
f_{ejecta,i}={\sum_{k>k_{max}}g_{ik} \Delta m_k \over \sum_j
g_{ji} \Delta m_j}, \label{fejecta_equation}
\end{equation}
where $k_{max}$ corresponds to the shell with the highest entropy fluid still gravitationally bound to the remnant.  In equation (\ref{fejecta_equation}), the sum over shells $k$ includes only those shells with larger post-shock entropies than this maximum, that is, only those shells associated with ejecta.
Figure \ref{fejecta} displays the $f_{ejecta}$ curves for the parent stars in a variety of different collision scenarios, both as determined by this method and as determined by an SPH calculation.

Lithium is a particularly interesting element to follow during a collision.
Lithium is burned during stellar
evolution except at low temperatures, and therefore can be used as an
indicator of mixing.  If a star has a deep enough surface convective
layer, there will be essentially no lithium, because the convection
mixes any lithium from the outer layers into the hot interior where it
is burned.  A small amount of lithium does exist in the outer few
percent of, for example, a $0.8 M_\odot$ turnoff star
(see Fig.\ \ref{chempar}) and would consequently become part of
the ejecta during a collision, resulting in a remnant with very little lithium.

\subsubsection{Structure and Composition\label{structure}}

\begin{figure}
\plotone{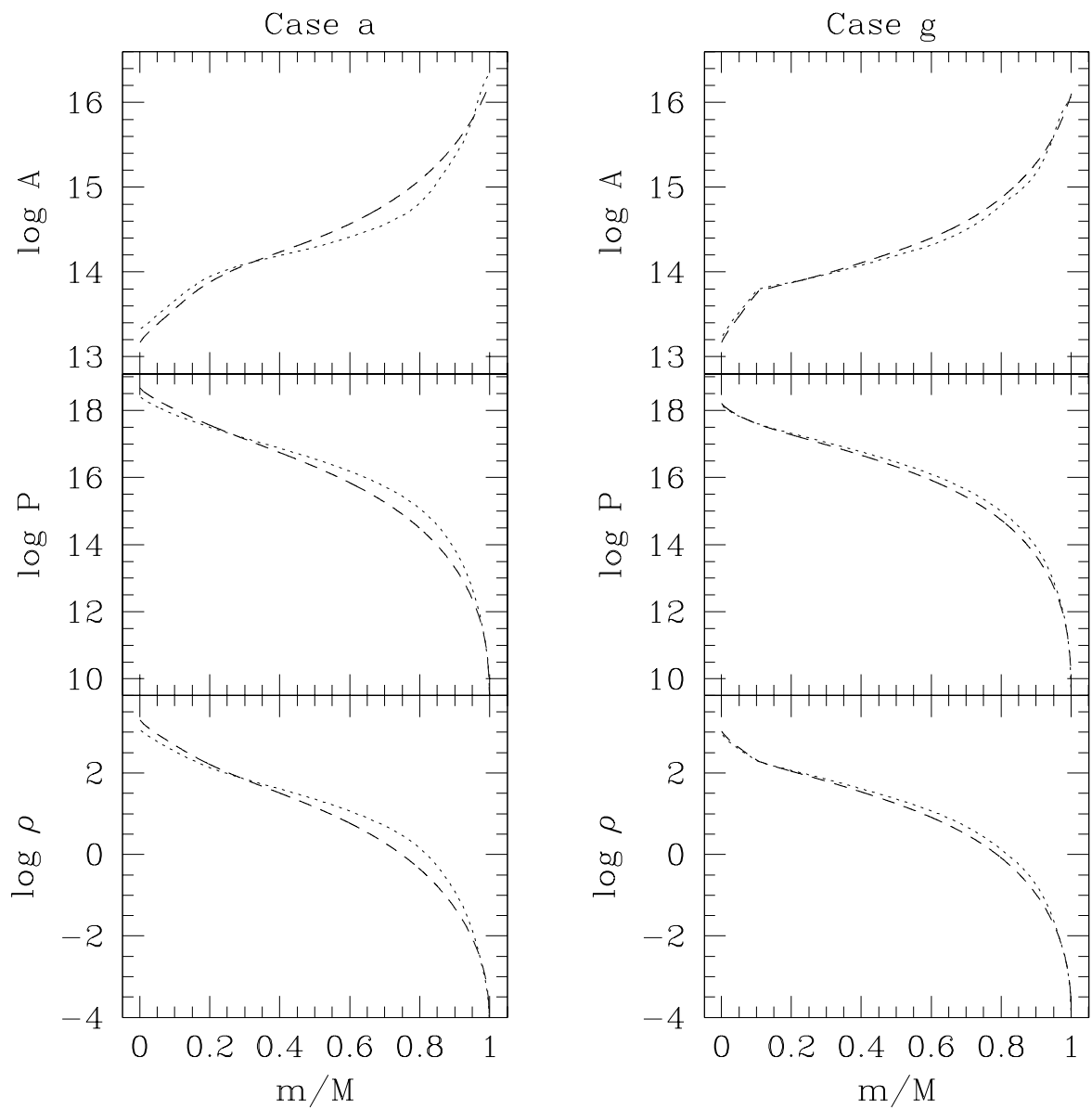}
%\plotone{thermag.eps}
\caption{Thermodynamic profiles of $A$, pressure
$P$, and density $\rho$ as a function of enclosed mass fraction
$m/M_r$ for the remnants of Cases a and g, where $M_r$ is the total
bound mass of the remnant.  The dotted line refers to the remnant
resulting from a 3D SPH simulation, and the dashed line refers to
the remnant generated by the method of this paper. Logarithms are
base 10 and units are cgs. \label{thermag} }
\end{figure}

\begin{figure}
\plotone{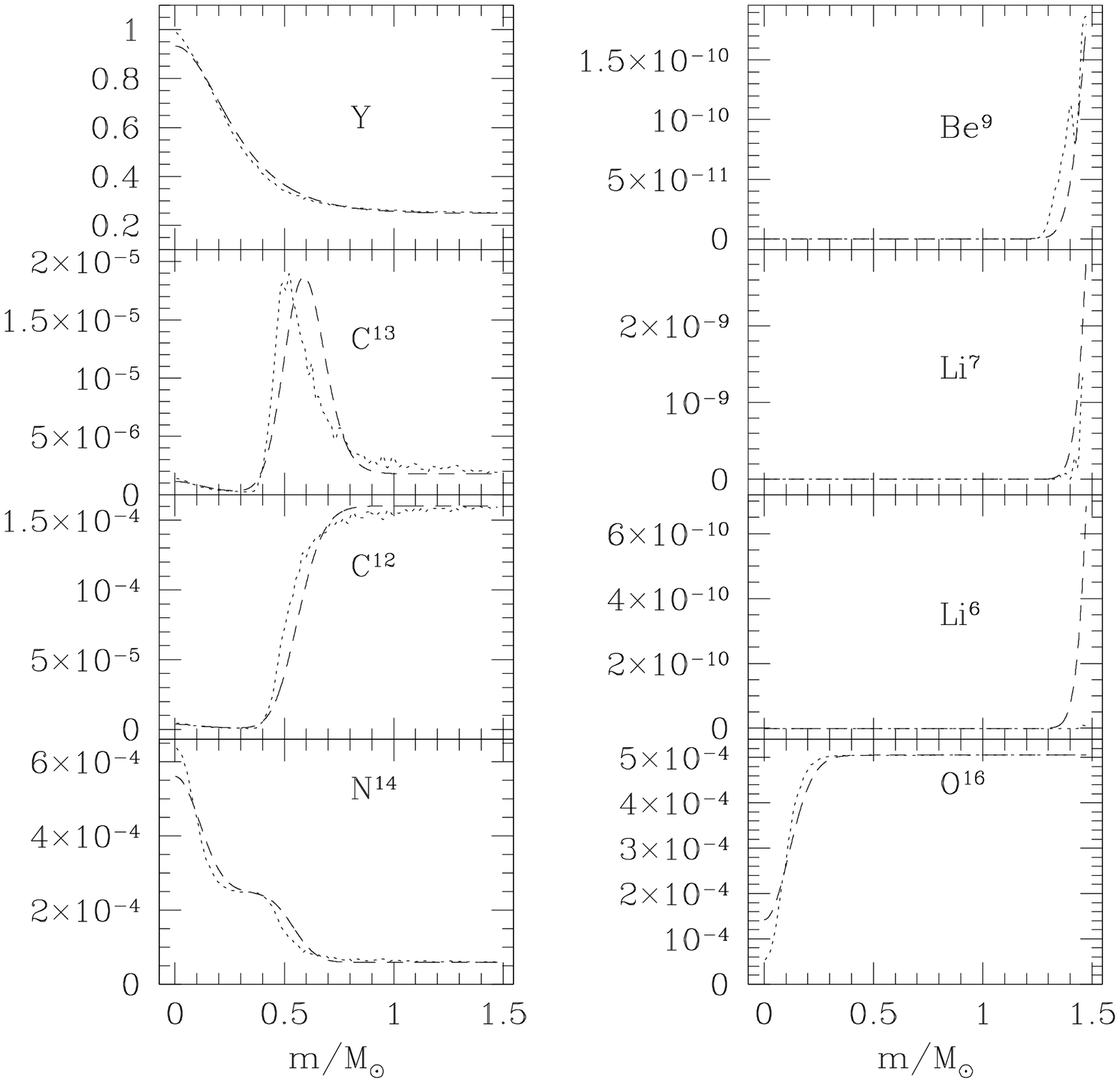}
%\plotone{chem8amdl.eps}
\caption{Fractional chemical abundance (by mass)
as a function of enclosed mass fraction $m/M_r$ for the Case a
remnant.  Line types are as in Fig.\ \ref{thermag}. \label{chem8a}
}
\end{figure}

\begin{figure}
\plotone{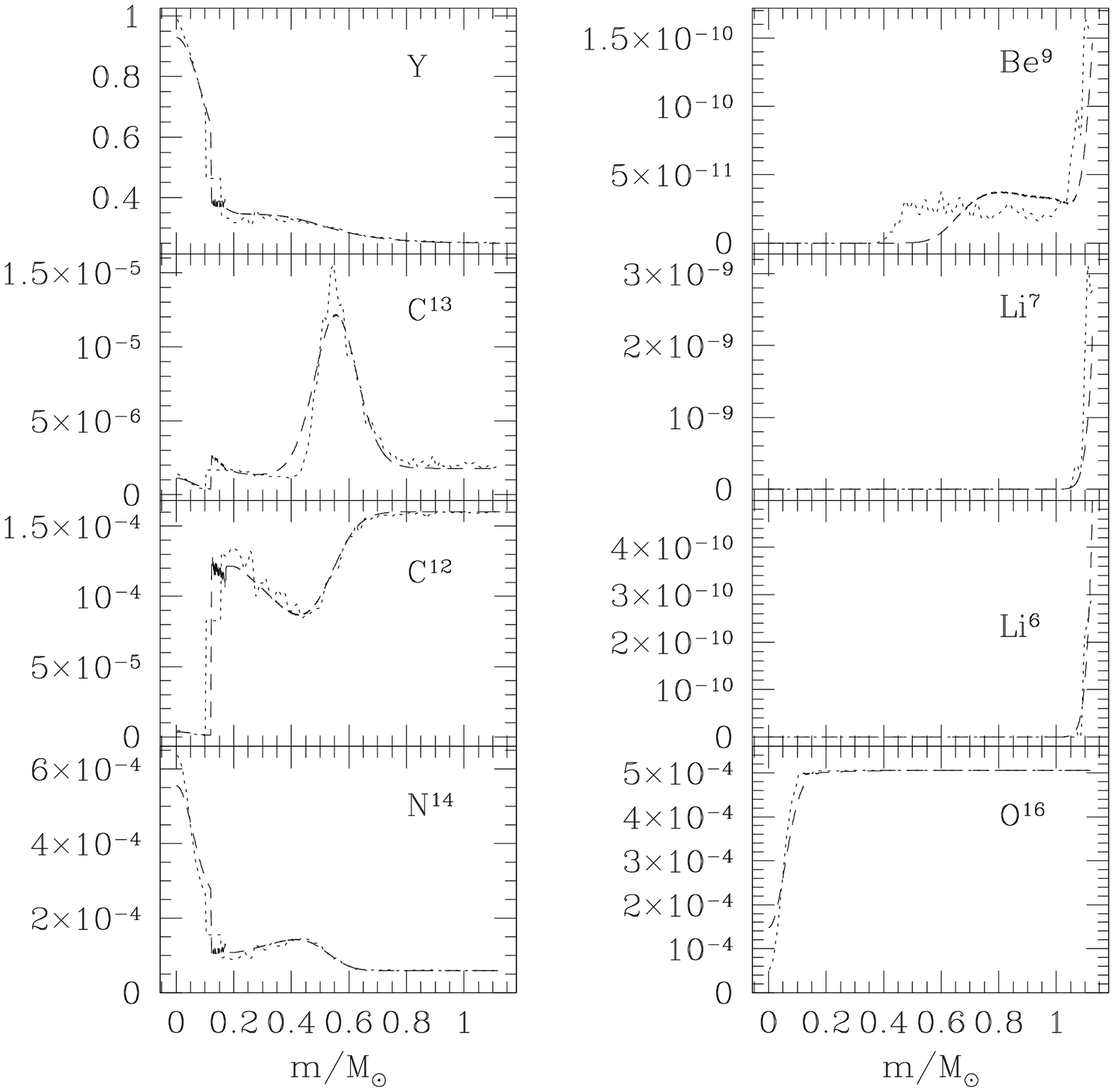}
%\plotone{chem8gmdl.eps}
\caption{Fractional chemical abundance (by mass)
as a function of enclosed mass fraction $m/M_r$ for the Case g
remnant.  Line types are as in Fig.\ \ref{thermag}. \label{chem8g}
}
\end{figure}

\begin{figure}
\plotone{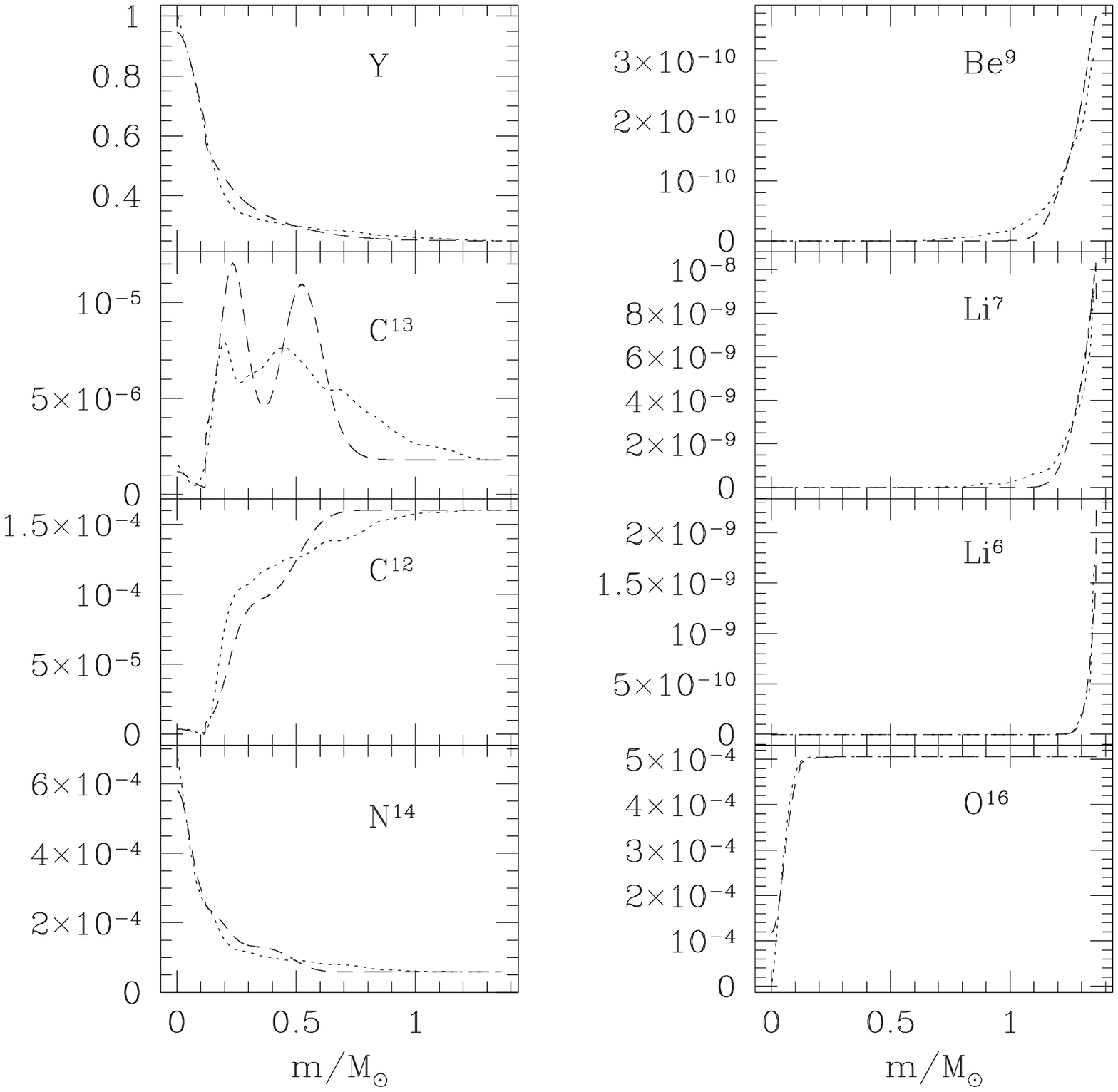}
%\plotone{chem8emdl.eps}
\caption{Fractional chemical abundance (by mass)
as a function of enclosed mass fraction $m/M_r$ for the Case e
remnant.  Line types are as in Fig.\ \ref{thermag}. \label{chem8e}
}
\end{figure}

\begin{figure}
\plotone{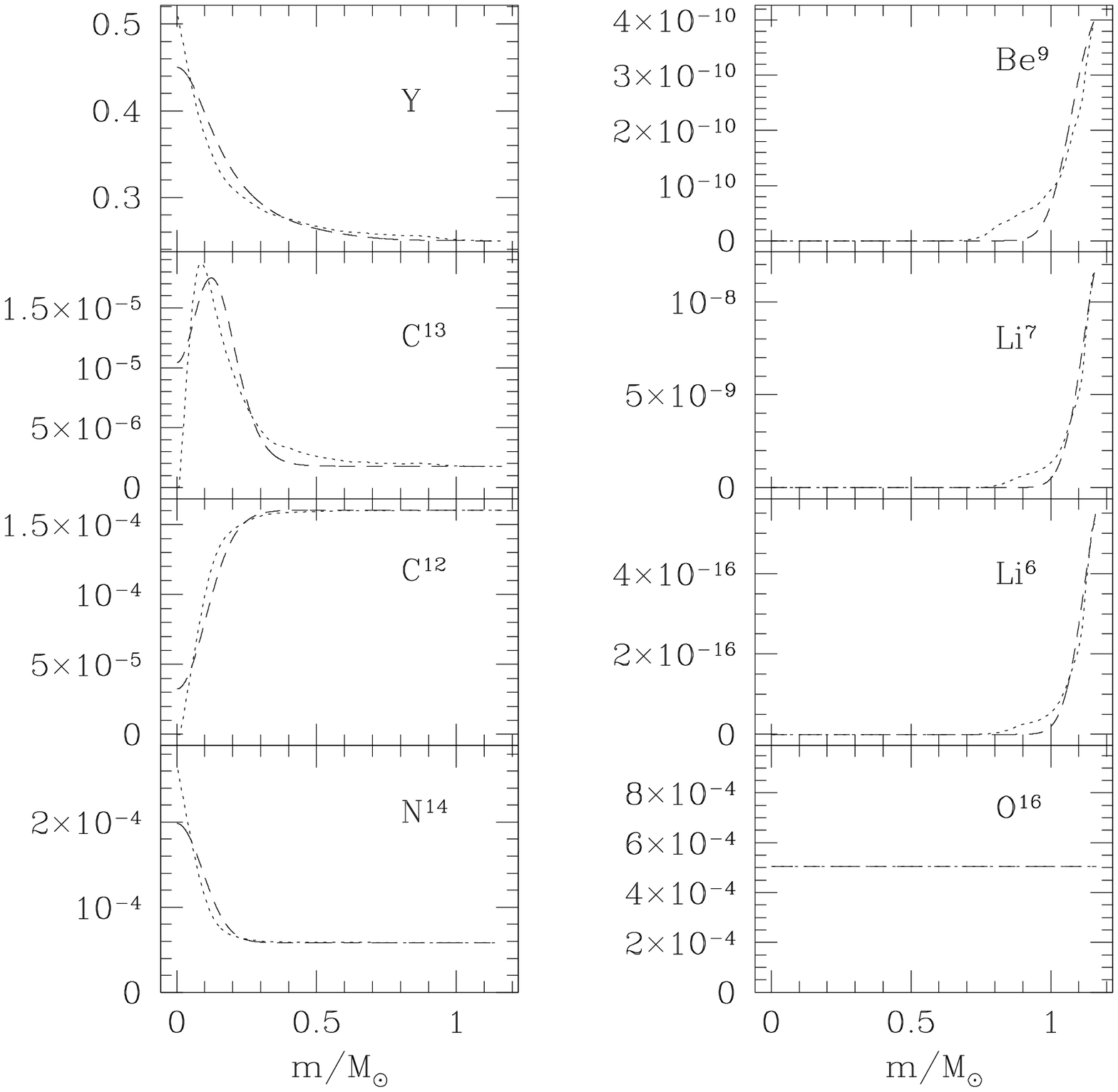}
%\plotone{chem8kmdl.eps}
\caption{Fractional chemical abundance (by mass)
as a function of enclosed mass fraction $m/M_r$ for the Case k
remnant.  Line types are as in Fig.\ \ref{thermag}. \label{chem8k}
}
\end{figure}

Thermodynamic (Fig.\ \ref{thermag}) and chemical composition (Figs.\
\ref{chem8a}, \ref{chem8g}, \ref{chem8e} and \ref{chem8k}) profiles
show that our remnant models are quite accurate.  In Case~g, our
remnant displays the kink in the $A$ profile near $m/M=0.1$ (see Fig.\
\ref{thermag}), inside of which the fluid originates solely from the
$0.8M_\odot$ star.  Our models also reproduce the chemical profiles of
the SPH remnant very well: the peak values in the chemical abundances
are often accurate to within 20\%, and the shapes of these profiles,
though sometimes peculiar, are followed closely.  Helium distribution
is particularly important to model well since it will help determine the MS
lifetime of the remnant.  As mentioned in \S\ref{results:shock}, the core of the remnant is usually somewhat overmixed, flattening out the helium profile in that region.  Nevertheless, the central value of the fractional helium
abundance $Y$ given by our models typically underestimates the SPH result
by only about 5\%.

Although near the remnant's
surface our method sometimes yields a large {\it fractional\/} error in
lithium abundance (see Figs.\ \ref{chem8a} and \ref{chem8g}),
this is simply because the overall abundance is so close to
zero.  For example, the predicted surface fractional Li$^6$
abundance of $6.9 \times 10^{-10}$ for our Case~a remnant is an
overestimate, but is nearly 20 times smaller than the surface
fractional abundance in the $0.8 M_\odot$ parent. Except for in
the extreme case of grazing collisions (when mass loss is
exceedingly small), collisional blue stragglers should be
severely depleted in lithium, a prediction that can be tested
with appropriate observations [see \citet{ss00} and
\citet{rbkr01}].

\subsubsection{Angular Momentum Distribution}

Our previous hydrodynamic simulations involving polytropic stars \citep{lom96} implemented
the classical form of the artificial viscosity (AV), which introduces a
significant amount of spurious shear in our differentially
rotating remnants.
The effects of shear are discussed in \S 4.2 of \citet{lom96} and
studied in detail in \citet{lom99}.
Shear tends to weaken differential
rotation, transporting angular momentum outward.  This angular
momentum transport acts on the viscous timescale, which
is comparable to the total time of a typical
simulation in our collisions between polytropic stars.
We therefore avoid comparisons involving the angular
momentum profiles of remnants from our polytropic SPH calculations.

Our collisions between realistically modeled stars
implemented the Balsara AV \citep{bal95}, with a
viscous timescale that is significantly larger.
In particular, \citet{lom99} show that the
viscous timescale
scales approximately with $N_N^{1/2}$ for the classical AV
(where $N_N$ is the neighbor number) and $N_N$ for Balsara AV.
Since we used $N_N=64$ in our polytropic simulations and $N_N=100$ in our realistic simulations, the viscous timescale is
larger in our collisions involving realistic parent stars by
a factor of approximately $100/64^{1/2}\sim 10$.  Consequently, in our
SPH calculations done with the Balsara AV, only a
relatively small amount of specific angular momentum is spuriously transported
out from the core.

\begin{figure}
\plotone{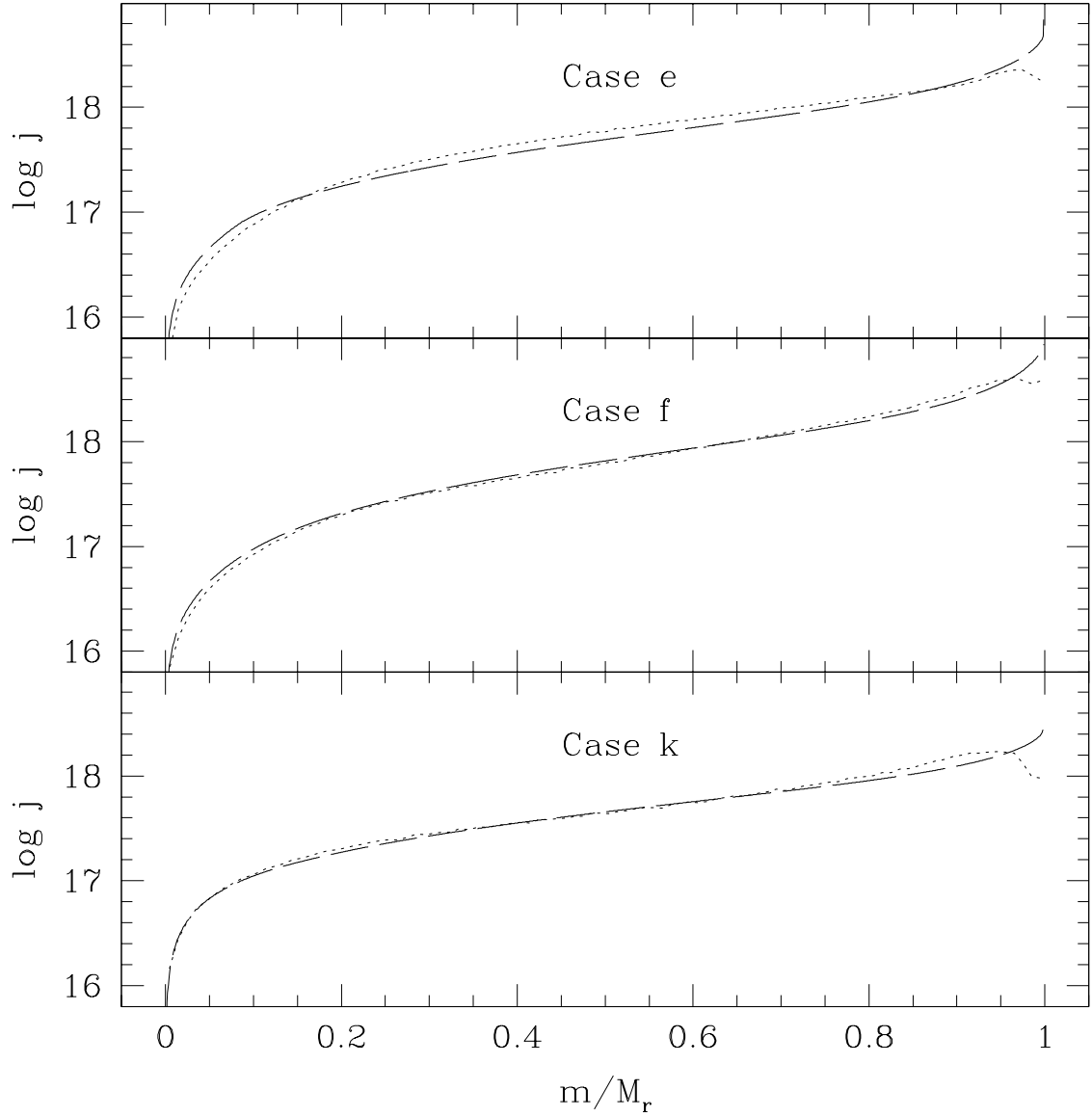}
%\plotone{log3mdl.eps}
\caption{SPH specific angular momentum profiles
(dotted curves) compared with the approximate profiles (long dashed curves) generated
from equation (\ref{jofm}).
Logarithms are base 10 and units are cgs. \label{jrealmdl} }
\end{figure}

Specific angular momentum profiles, averaged over surfaces of
constant density, are compared in Figure \ref{jrealmdl} for the
three realistically modeled cases with rotating remnants: Cases
e, f and k. We find excellent agreement between our simple models
and their corresponding SPH counterparts.  Our procedure for
determining the angular momentum distribution, as described in \S
\ref{angular}, yields values of $k_1$, $k_2$ and $k_3$ of,
respectively, 0.161, 0.427 and $3.16\times 10^{16}$ cm$^2$/s for
Case e, of 0.023, 0.570 and $5.93\times 10^{15}$ cm$^2$/s for Case
f, and of 0.146, 0.415 and $4.56\times 10^{16}$ cm$^2$/s for Case
k. Note that the hump in the SPH $j$ profile in the outer few
percent of the remnant results from the need to terminate the
simulation before all of the gravitationally bound fluid has
fallen back to the merger remnant: this artifact is gradually
diminishing during the final stages of the SPH calculation.

\subsection{Stellar Evolution of Remnant Models}

A rigorous test of the validity of the simple models, which we have
performed using YREC, is to compare their subsequent stellar evolution
against that of SPH-generated models.  YREC evolves a star through a
sequence of models of increasing age, solving the stellar evolution
equations for interior profiles such as chemical composition,
pressure, temperature, density and luminosity.  All relevant nuclear
reactions (including pp-chains, the CNO cycle, triple-$\alpha$
reactions and light element reactions) are treated. Recent opacity
tables are used (ensuring that the remnant's position in a CMD can be
accurately determined) and mixing mechanisms are incorporated. For
blue stragglers, the various mixing processes can potentially carry
fresh hydrogen fuel into the stellar core and thereby extend the MS
lifetime of the remnant. Furthermore, any helium mixed into the outer
layers affects the opacity and hence the remnant's position in a
CMD. The free parameters in YREC (e.g., the mixing length) are set by
calibrating a solar mass and solar metallicity model to the Sun.

\begin{figure}
\plotone{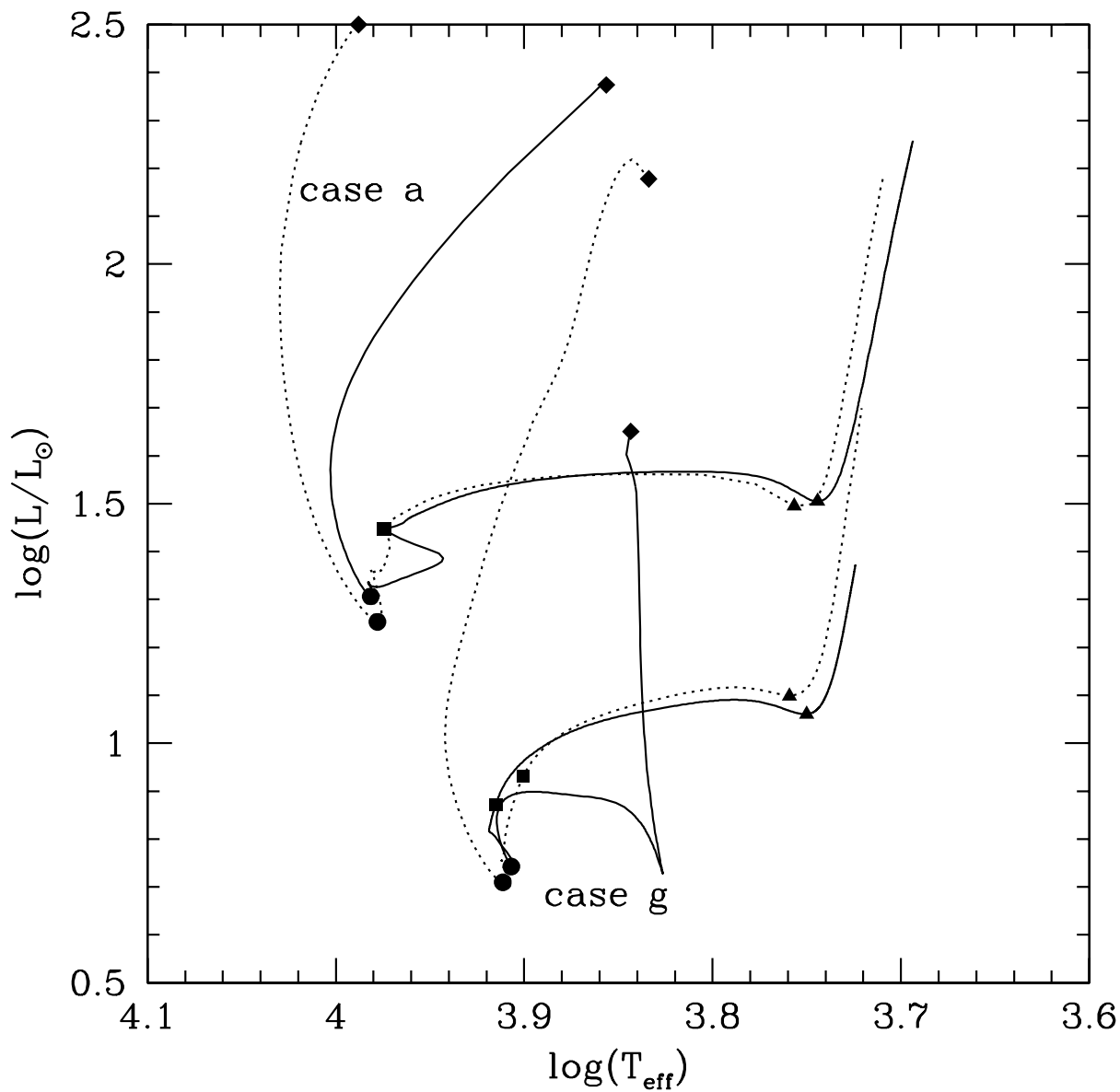}
%\plotone{evolution.eps}
\caption {Evolutionary tracks for Case a and
Case g. The solid lines correspond to the tracks of the simple models
described in this paper, while the dotted lines show tracks for
models generated directly from the output of SPH
calculations. The symbols mark identifiable evolutionary phases
in the evolution of the collision products: diamonds show the position
of the star immediately following the collision. Circles mark the
``zero age'' MS, squares mark the MS turnoff,
and triangles show the base of the giant branch. \label{evolve_ag}}
\end{figure}

Using the method described in \citet{sil97b}, we used two of our
simple models (Cases a and g) as starting models in YREC and evolved
the collision products from the end of the collision to the giant
branch.  Figure \ref{evolve_ag} shows the evolutionary tracks for
these simple models (solid lines) and the SPH-generated models (dotted
lines).  The tracks of the SPH-generated models are discussed in
\citet{sil97}.

The agreement between the two sets of models is very good.  Although
there are some differences on the 'pre-MS' portion of the tracks, this
stage only lasts for approximately a Kelvin-Helmholtz timescale, a
very small fraction of the total lifetime of the cluster.  Indeed, for
Case a, the SPH and simple models reach the MS after 0.4 Myr and 0.8
Myr, respectively, and the corresponding contraction times for the
Case g models are only 4 Myr and 2 Myr.  Since the stars contract to
the MS so quickly, the exact pre-MS track may not be directly
important for generating synthetic CMDs.  Nevertheless a reasonable
model of a pre-MS star can be of interest: the radius of a newly born
remnant, and hence its collisional cross-section in a multi-star
interaction, is strongly dependent on how the fluid has been shocked,
and furthermore the surface abundances are strongly dependent on how
it has been mixed.

Once the collision product reaches the MS, the two methods show very
good agreement, and the subgiant and giant branch evolution of these
stars is virtually identical.  The MS lifetimes for the two different
methods agree reasonably well. For Case g, the SPH results give a
lifetime of 850 Myr, while the simple models give 650 Myr. For Case a,
the SPH results give a lifetime of 80 Myr, compared to 180 Myr from
the simple models. It should be noted that the Case a remnant has
central helium abundance near 100\%, accounting for its short MS
lifetime.  For such remnants, even a slight inaccuracy in the core's
helium profile has a large relative effect on how long the star
remains on the MS.  While the MS lifetime resulting from our method
can be off by more than a factor of two for remnants with intact
helium cores, the lifetime of such remnants is nevertheless a very
small fraction of the lifetime of globular clusters, and therefore the
simple models can still be useful for incorporating stellar collisions
into dynamical models of globular cluster evolution.

\section{Concluding Remarks}

An important question in the study of globular clusters is what
are the necessary features of stellar collisions that must be
modeled in order to synthesize reliable theoretical CMDs. While
detailed studies of the tracks and evolutionary timescales of
rapidly rotating collision remnants will be necessary to answer
this question fully, we do feel that the features modeled in this
paper (mass loss, shock heating, hydrodynamic mixing, angular
momentum distribution) are all essential components to consider.
Our treatment of ejecta allows for an accurate estimate of the
total mass of the remnant, upon which the subsequent stellar
evolution and MS lifetime depend sensitively. Furthermore, shock
heating during a collision not only influences the structure of
the remnant, but also helps determine if convective regions can
develop during the future evolution: the entropy gradients implied
by SPH results and by the shock heating method of this paper tend
to stabilize a contracting pre-MS star against convection.
Previous stellar evolution studies of remnants have found that the
tracks are sensitive to the assumptions made about how the fluid
is mixed during a merger \citep{bai95b,sil99}; our simple fluid
mixing algorithms give a compromise between previously used
approximations that tend to bracket the actual amount of mixing
during a collision.  Finally, although the treatment of rotation
in stellar evolution is a challenging problem, it is clear that
rotational support and induced mixing must be considered as they
have profound consequences on stellar evolution
\citep{sil00b,sil01}; the form for the angular momentum
distribution presented in this paper provides a simple means of
generating very reasonable initial profiles for future studies of
collisional remnants.

The algorithms we have developed are implemented in a publicly
available FORTRAN software package named ``Make Me A Star.''  For the
forseeable future, this package can be downloaded from
http://vassun.vassar.edu/$\sim$lombardi/mmas/.
Researchers
should be aware of the limitations of this method in terms of
the structure, rotational properties and evolutionary timescales of
the models created, as outlined in this paper.
This software does produce accurate models for a variety of collision
scenarios,
and we hope that it will be used in combination with realistic dynamical simulations of
star clusters that must take into account stellar collisions.

\acknowledgments

We thank J.\ Faber for his contributions to the SPH code, A.\
Thrall for testing our software package and for helpful comments,
and the anonymous referee for valuable comments.
J.C.L.\ acknowledges support from the Keck Northeast Astronomy
Consortium, from a grant from the Research Corporation, and from
NSF Grant AST-0071165. F.A.R.\ acknowledges support from NSF
Grants AST-9618116 and PHY-0070918, NASA ATP Grant NAG5-8460, and
a Sloan Research Fellowship. This work was also partially
supported by the National Computational Science Alliance under
Grant AST980014N and utilized the NCSA SGI/Cray Origin2000
parallel supercomputer.

\clearpage

\input tab1.tex
\input tab2.tex
\input tab3.tex
\input tab4.tex

\end{document}

%% file: tab1.tex
\begin{deluxetable}{clccc}
%\tabletypesize{\scriptsize}
\tablecaption{Parent Star Characteristics \label{tbl-1}}
\tablewidth{0pt} \tablehead{ Structure type & \colhead{$M [M_\odot]$}
& \colhead{$R [R_\odot]$} & \colhead{$R_{0.5} [R_\odot]$} &
\colhead{$R_{0.86} [R_\odot]$} } \startdata
$n=3$ polytrope     & 0.8   & 0.955  & 0.270 & 0.443\\
Composite polytrope & 0.6   & 0.535  & 0.253 & 0.382\\
$n=1.5$ polytrope   & 0.4   & 0.353  & 0.184 & 0.261\\
$n=1.5$ polytrope   & 0.16  & 0.153  & 0.080 & 0.113\\
YREC                & 0.8   & 0.955  & 0.198 & 0.395\\
YREC                & 0.6   & 0.517  & 0.203 & 0.332\\
YREC                & 0.4   & 0.357  & 0.182 & 0.272\\
 \enddata
\end{deluxetable}

%% file: tab2.tex
\begin{deluxetable}{ccccllcccc}
%\tabletypesize{\scriptsize}
\tablecaption{Mass loss \label{tbl-2}} \tablewidth{0pt}
\tablehead{ \colhead{Case\tablenotemark{a}} & \colhead{${M_1
\over M_\odot}$}   & \colhead{${M_2\over M_\odot}$}   &
\colhead{${r_p\over R_1+R_2}$}  & \colhead{$f_{2,SPH}$\tablenotemark{b}} & \colhead{$f_2$\tablenotemark{c}} &
\colhead{$f_{L,SPH}$\tablenotemark{d}}  &
\colhead{$f_L$\tablenotemark{e}} & \colhead{${M_{SPH}\over
M_\odot}$\tablenotemark{f}}  & \colhead{${M_r\over 
M_\odot}$\tablenotemark{g}}} \startdata
%  M_1    M_2   r_p    f_2,SPH  f_2  f_L,SPH   f_L   M_SPH   M_r
A & 0.8 & 0.8 & 0.00  & 0.50 & 0.50 & 0.064 & 0.064 & 1.50 & 1.497\\
B & 0.8 & 0.8 & 0.25  & 0.51 & 0.50 & 0.023 & 0.025 & 1.56 & 1.560\\
C & 0.8 & 0.8 & 0.50  & 0.50 & 0.50 & 0.012 & 0.015 & 1.58 & 1.577\\
D & 0.8 & 0.6 & 0.00  & 0.24 & 0.24 & 0.057 & 0.061 & 1.32 & 1.315\\
E & 0.8 & 0.6 & 0.25  & 0.19 & 0.26 & 0.024 & 0.027 & 1.37 & 1.363\\
F & 0.8 & 0.6 & 0.50  & 0.26 & 0.30 & 0.008 & 0.017 & 1.39 & 1.376\\
G & 0.8 & 0.4 & 0.00  & 0.049& 0.042& 0.056 & 0.054 & 1.13 & 1.135\\
H & 0.8 & 0.4 & 0.25  & 0.059& 0.065& 0.028 & 0.024 & 1.17 & 1.172\\
I & 0.8 & 0.4 & 0.50  & 0.11 & 0.12 & 0.008 & 0.015 & 1.19 & 1.182\\
J & 0.6 & 0.6 & 0.00  & 0.50 & 0.50 & 0.049 & 0.059 & 1.14 & 1.129\\
K & 0.6 & 0.6 & 0.25  & 0.50 & 0.50 & 0.028 & 0.030 & 1.17 & 1.164\\
L & 0.6 & 0.6 & 0.50  & 0.51 & 0.50 & 0.022 & 0.020 & 1.17 & 1.175\\
M & 0.6 & 0.4 & 0.00  & 0.11 & 0.13 & 0.054 & 0.055 & 0.95 & 0.945\\
N & 0.6 & 0.4 & 0.25  & 0.14 & 0.18 & 0.029 & 0.029 & 0.97 & 0.971\\
O & 0.6 & 0.4 & 0.50  & 0.16 & 0.26 & 0.010 & 0.020 & 0.99 & 0.980\\
P & 0.4 & 0.4 & 0.00  & 0.50 & 0.50 & 0.037 & 0.056 & 0.77 & 0.756\\
Q & 0.4 & 0.4 & 0.25  & 0.51 & 0.50 & 0.029 & 0.030 & 0.78 & 0.776\\
R & 0.4 & 0.4 & 0.50  & 0.54 & 0.50 & 0.010 & 0.020 & 0.79 & 0.784\\
S & 0.4 & 0.4 & 0.75  & 0.47 & 0.50 & 0.008 & 0.015 & 0.79 & 0.788\\
T & 0.4 & 0.4 & 0.95  & 0.51 & 0.50 & 0.011 & 0.013 & 0.79 & 0.790\\
U & 0.8 & 0.16& 0.00  & 0.015& 0.001& 0.027 & 0.035 & 0.93 & 0.927\\
V & 0.8 & 0.16& 0.25  & 0.016& 0.003& 0.025 & 0.014 & 0.94 & 0.946\\
W & 0.8 & 0.16& 0.50  & 0.024& 0.014& 0.021 & 0.009 & 0.94 & 0.951\\
a & 0.8 & 0.8 & 0.00  & 0.50 & 0.50 & 0.079 & 0.078 & 1.47 & 1.475\\
e & 0.8 & 0.6 & 0.25  & 0.19 & 0.20 & 0.029 & 0.026 & 1.36 & 1.363\\
f & 0.8 & 0.6 & 0.50  & 0.21 & 0.21 & 0.014 & 0.016 & 1.38 & 1.377\\
g & 0.8 & 0.4 & 0.00  & 0.098& 0.039& 0.062 & 0.061 & 1.13 & 1.127\\
k & 0.6 & 0.6 & 0.25  & 0.50 & 0.50 & 0.032 & 0.030 & 1.16 & 1.164\\
 \enddata

%% Text for table notes should follow after the \enddata but before
%% the \end{deluxetable}. Make sure there is at least one \tablenotemark
%% in the table for each \tablenotetext.

\tablenotetext{a}{Capital letters refer to collisions of
polytropic stars; lower case letters to that
of more realistically modeled parents}
\tablenotetext{b}{Fraction of ejecta originating in star 2, as
determined by SPH}
\tablenotetext{c}{Fraction of ejecta originating in star 2,
as determined by method of this paper}
\tablenotetext{d}{Total fractional mass loss, as
determined by SPH}
\tablenotetext{e}{Total fractional
mass loss, as estimated by equation (\ref{fL})}
\tablenotetext{f}{Remnant mass, as determined by SPH}
\tablenotetext{g}{Remnant mass, as estimated by
$(1-f_L)(M_1+M_2)$}

%\tablecomments{Occasionally, authors wish to append a short
%paragraph of explanatory notes that pertain to the entire table, but
%which are different than the caption.  Such notes should be placed in
%a {\tt tablecomments} command like this.}

\end{deluxetable}

%% file: tab3.tex
\begin{deluxetable}{cccc}
%\tabletypesize{\scriptsize}
\tablecaption{Total Energy\label{tbl-energy}}
\tablewidth{0pt}
\tablehead{
\colhead{Case} &
\colhead{$E_{tot}$\tablenotemark{a}} &
\colhead{$E_{r,SPH}$\tablenotemark{b}}  &
\colhead{$E_r$\tablenotemark{c}}\\
 &
\colhead{[$10^{48}$g cm$^2$/s$^2$]} &
\colhead{[$10^{48}$g cm$^2$/s$^2$]}    &
\colhead{[$10^{48}$g cm$^2$/s$^2$]}
}
\startdata
A &-3.81&-4.5 & -4.43\\
D &-3.09&-3.6 & -3.56\\
G &-2.64&-3.1 & -3.00\\
J &-2.37&-2.6 & -2.72\\
M &-1.92&-2.2 & -2.19\\
P &-1.47&-1.6 & -1.68\\
U &-2.18&-2.3 & -2.37\\
a &-5.23&-6.3 & -6.25\\
g &-3.35&-3.9 & -3.86\\
\enddata

%\tablecomments{}

\tablenotetext{a}{The total energy in the system's
center of mass frame.} \tablenotetext{b}{The energy of
the remnant in its center of mass frame, as determined by an SPH
simulation} \tablenotetext{c}{The energy of the remnant
in its center of mass frame, as determined by eq.\ (\ref{Etot})}

\end{deluxetable}

%% file: tab4.tex
\begin{deluxetable}{cccc}
%\tabletypesize{\scriptsize}
\tablecaption{Total Angular Momentum  \label{tbl-angular_momentum}}
\tablewidth{0pt}
\tablehead{
\colhead{Case} &
\colhead{$J_{tot}$\tablenotemark{a}} &
\colhead{$J_{r,SPH}$\tablenotemark{b}}  &
\colhead{$J_r$\tablenotemark{c}}\\
 &
\colhead{[$10^{51}$g cm$^2$/s]} &
\colhead{[$10^{51}$g cm$^2$/s]}    &
\colhead{[$10^{51}$g cm$^2$/s]}
}
\startdata
B & 2.99& 2.84& 2.84\\
C & 4.23& 4.13& 4.10\\
E & 2.12& 2.04& 2.00\\
F & 2.99& 2.93& 2.89\\
H & 1.43& 1.34& 1.36\\
I & 2.02& 1.97& 1.96\\
N & 0.97& 0.91& 0.91\\
O & 1.37& 1.33& 1.31\\
Q & 0.64& 0.61& 0.60\\
R & 0.91& 0.90& 0.87\\
S & 1.11& 1.10& 1.08\\
T & 1.25& 1.21& 1.22\\
V & 0.59& 0.54& 0.57\\
W & 0.83& 0.75& 0.82\\
e & 2.10& 1.99& 1.99\\
f & 2.97& 2.85& 2.87\\
k & 1.43& 1.36& 1.34
\enddata
\tablenotetext{a}{The total angular momentum in the system's
center of mass frame.} \tablenotetext{b}{The angular momentum of
the remnant in its center of mass frame, as determined by an SPH
simulation} \tablenotetext{c}{The angular momentum of the remnant
in its center of mass frame, as determined by eq.\ (\ref{Jtot})}

\end{deluxetable}

%% file: ms.bbl
\begin{thebibliography}{}
%\bibitem{BH} Benz, W. \& Hills, J. G. 1987, \apj, 323, 614
%\bibitem{GH} Goodmann, J., \& Hernquist, L. 1991, \apj, 378, 637
%\bibitem{yrec} Guenther, D. B., Demarque, P., Kim, Y.-C., \& Pinsonneault, M.\ H., 1992 \apj, 387, 372
%\bibitem{Huang} Huang, K. 1987, Statistical Mechanics, (New York: John Wiley and Sons)
%\bibitem{LRS2}  Lombardi, J. C., Rasio, F. A., \& Shapiro, S.L. 1996, \apj, 468, 797
%\bibitem{} Rasio, F. A. 1991, PhD Thesis, Cornell University
%\bibitem{sand} Sandquist, E., Bolte, M., \& Hernquist, L. 1997, \apj, 477, 335
%\bibitem{Sills2} Sills, A., Lombardi, J. C., Bailyn, C. D., Demarque, P., Rasio, F. A., \& Shapiro, S. L. 1997 \apj, in press
\bibitem[Bacon et al.(1996)]{bac96} Bacon, D., Sigurdsson, S., \& Davies, M.\ B. 1996, \mnras, 281, 830
\bibitem[Balsara(1995)]{bal95} Balsara, D. 1995, J.~Comput.~Phys., 121, 357
\bibitem[Bailyn(1995)]{bai95} Bailyn, C.\ D.\ 1995, Ann. Rev. Ast. Astrop., 33, 133
\bibitem[Bailyn \& Pinsonneault(1995)]{bai95b} Bailyn, C.\ D.\ \&
Pinsonneault, M.\ H.\ 1995, \apj, 439, 705
\bibitem[Davies \& Benz(1995)]{dav95} Davies, M.\ B., \& Benz, W.\ 1995, \mnras, 276, 876
\bibitem[Endal \& Sofia(1976)]{ES76} Endal, A. S., \& Sofia, S. 1976,
\apj, 210, 184
\bibitem[Gilliland et al.(1998)]{gil98} Gilliland, R.\ L.\ 1998, \apj, 507, 818
\bibitem[Hills \& Day(1976)]{hil76} Hills, J.\ G., \& Day, C.\ A. 1976, Ap. Letters, 17, 87
\bibitem[Hut et al.(1992)]{hut92}Hut, P., McMillan, S., Goodman, J., Mateo, M., Phinney, E.\ S., Pryor, C., Richer, H.\ B., Verbunt, F.\ \& Weinberg, M.\ 1992, \pasp, 104, 981
\bibitem[Lai, Rasio \& Shapiro(1994)]{lai94} Lai, D.\, Rasio, F.\ A.\ \& Shapiro, S.\ L.\ 1994, \apj, 423, 344.
\bibitem[Leonard(1989)]{leo89} Leonard, P.\ J.\ T.\ 1989, \aj, 98, 217
\bibitem[Leonard \& Fahlman(1991)]{leo91} Leonard, P.\ J.\ T., \& Fahlman, G.\ G.\ 1991, \aj, 102, 994
\bibitem[Livio(1993)]{liv93} Livio, M.\ 1993, in {\it Blue Stragglers\/}, ASP Conf.\ Ser.\ 53, ed.\ R.\ A.\ Saffer (San Francisco: ASP)
\bibitem[Lombardi et al.(1996)]{lom96} Lombardi, J.\ C., Rasio, F.\ A., \& Shapiro, S.\ L.\ 1996, \apj, 468, 797
\bibitem[Lombardi et al.(1999)]{lom99} Lombardi, J.\ C., Sills, A., Rasio, F.\ A., \& Shapiro, S.\ L.\ 1999,
 J.\ Comp.\ Phys., 152, 687
\bibitem[Monaghan(1992)]{mon92} Monaghan, J.\ J.\ 1992, \araa, 30, 543
%\bibitem[Ouellette \& Pritchet(1998)]{oue98} Ouellette, J.\ A.\ \& Pritchet,
%C.\ J.\ 1998, \aj, 115, 2539
\bibitem[Rasio \& Lombardi(1999)]{ras99} Rasio, F.\ A., \& Lombardi, J.\ C.\ 1999, J.\ Comp.\ App.\ Math.,
109, 213
\bibitem[Rasio, Fregeau \& Joshi (2001)]{ras01} Rasio, F.\ A., Fregeau, J.\ M., \& Joshi, K.\ J.\ 2001, to appear in
{\it The Influence of Binaries on Stellar Population Studies\/}, 
ed. D.~Vanbeveren (Dordrecht: Kluwer). [astro-ph/0103001]
\bibitem[Ryan et al.(2001)]{rbkr01}Ryan, S., Beers, T., Kajino, T.\ \& Rosolankova, K.\ 2001, \apj, 547, 231
\bibitem[Sandquist et al.(1997)]{san97} Sandquist, E.\ L., Bolte, M.\ \& Hernquist, L.\ 1997, \apj, 477, 335
\bibitem[Sepinsky et al.(2000)]{sep00} Sepinsky, J.\ F., Saffer, R.\ A., Pilman, C.\ S., DeMarchi,G., \&
Paresce, F.\ 2000, AAS Meeting 196, \#41.06
\bibitem[Shara et al.(1997)]{sha97} Shara, M.\ M., Saffer, R.\ A., \& Livio, N.\ 1997, \apjl, 489, L59
\bibitem[Shetrone \& Sandquist(2000)]{ss00} Shetrone, M.\ D., \&
Sandquist, E.\ L. 2000, \aj, 120, 1913
\bibitem[Sigurdsson et al.(1994)]{sig94} Sigurdsson, S., Davies, M.\ B., \& Bolte, M.\ 1994, \apjl, 431, L115
\bibitem[Sigurdsson \& Phinney(1995)]{sig95} Sigurdsson, S., \& Phinney, E.\ S.\ 1995, \apjs, 99, 609
\bibitem[Sills \& Bailyn(1999)]{sil99} Sills, A., \& Bailyn, C.\ D.\ 1999, \apj, 513, 428
\bibitem[Sills \& Lombardi(1997)]{sil97} Sills, A.\ \& Lombardi, J.\ C.\ 1997, \apjl, 484, L51
\bibitem[Sills et al.(1997)]{sil97b} Sills, A., Lombardi, J.\ C., Jr., Demarque, P.\ D., Bailyn, C.\ D.,
Rasio, F.\ A., \& Shapiro, S.\ L. 1997, \apj, 487, 290
\bibitem[Sills et al.(2000)]{sil00}Sills, A., Bailyn, C.\ D., Edmonds, P.\ D., \& Gilliland, R.\ L. 2000, \apj,
535, 298
\bibitem[Sills et al.(2001)]{sil01} Sills, A., Faber, J.\ A., Lombardi, J.\ C., Rasio, F.\ A., \&
Warren, A.\ R. 2001, \apj, 548, 323
\bibitem[Sills, Pinsonneault, \& Terndrup(2000)]{sil00b} Sills, A.,
Pinsonneault, M.\ H., \& Terndrup, D.\ M.\ 2000, \apj, 534, 335
\bibitem[Stryker(1993)]{str93} Stryker, L.\ L.\ 1993, \pasp, 105, 1081
\bibitem[Tassoul(1978)]{tas78} Tassoul, J.\ 1978, Theory of Rotating Stars (Princeton: Princeton Univ.\ Press)
\bibitem[Tassoul(2000)]{tas00} Tassoul, J.\ 2000, Stellar Rotation (Cambridge: Cambridge Univ.\ Press)
\end{thebibliography}
